\documentclass[amsmath,amssymb,aps,prl,superscriptaddress,twocolumn,10pt]{revtex4-1} 

\usepackage{float}
\usepackage{stmaryrd}
\usepackage[usenames,dvipsnames]{color}
\usepackage{graphicx}
\usepackage{wrapfig}
\usepackage{xcolor}
\usepackage{cancel}
\usepackage{amsmath}
\usepackage{times}
\usepackage[latin1]{inputenc}
\usepackage{setspace} 
\usepackage[normalem]{ulem}
\usepackage[hypertexnames=false,plainpages=false,pdfpagelabels,colorlinks=true,linkcolor=blue,urlcolor=blue,citecolor=blue,pdfauthor={Simon Braun},pdftitle={Emergence of Coherence and the Dynamics of Quantum Phase Transitions},pdfdisplaydoctitle=true,pdfpagelayout=OneColumn,pdfduplex=DuplexFlipLongEdge]{hyperref} 

\newcommand{\vspacelarge}{0.6cm}
\newcommand{\vspacesmall}{0.1cm}

\newcommand{\ket}[1]{\vert #1 \rangle}

\def\d{\dagger}

\begin{document}

\singlespacing
\noindent
\title{Emergence of coherence and the dynamics of quantum phase transitions}

\newcommand{\lmu}{Fakultät für Physik, Ludwig-Maximilians-Universität München, Schellingstr.\ 4, 80799 Munich, Germany}
\newcommand{\mpq}{Max-Planck-Institut für Quantenoptik, Hans-Kopfermann-Str.\ 1, 85748 Garching, Germany}
\newcommand{\fu}{Freie Universität Berlin, Arnimallee 14, 14195 Berlin, Germany}
\newcommand{\madrid}{Instituto de Fisica Fundamental, CSIC, Serrano 113-B, 28006 Madrid, Spain}
\newcommand{\golm}{Max-Planck-Institut für Gravitationsphysik, Am Mühlenberg 1, 14476 Potsdam-Golm, Germany}
\newcommand{\mail}{E-mail: ulrich.schneider@lmu.de}

\noindent
\author{S.\ Braun}
\affiliation{\lmu}
\affiliation{\mpq}

\author{M.\ Friesdorf} 
\affiliation{\fu}

\author{S.\ S.\ Hodgman}
\affiliation{\lmu}
\affiliation{\mpq}

\author{M.\ Schreiber}
\affiliation{\lmu}
\affiliation{\mpq}

\author{J.\ P.\ Ronzheimer}
\affiliation{\lmu}
\affiliation{\mpq}

\author{A.\ Riera}
\affiliation{\fu}
\affiliation{\golm}

\author{M.\ del Rey}
\affiliation{\madrid}

\author{I.\ Bloch}
\affiliation{\lmu}
\affiliation{\mpq}

\author{J.\ Eisert}
\affiliation{\fu}

\author{U.\ Schneider}
\affiliation{\lmu}
\affiliation{\mpq}
\affiliation{\mail}

\begin{abstract}
  The dynamics of quantum phase transitions poses one of the most challenging problems in modern many-body physics. Here, we study a prototypical example in a clean and well-controlled ultracold atom setup by observing the emergence of coherence when crossing the Mott insulator to superfluid quantum phase transition. In the one-dimensional Bose-Hubbard model, we find perfect agreement between experimental observations and numerical simulations for the resulting coherence length. We thereby perform a largely certified analogue quantum simulation of this strongly correlated system reaching beyond the regime of free quasiparticles. Experimentally, we additionally explore the emergence of coherence in higher dimensions where no classical simulations are available, as well as for negative temperatures. For intermediate quench velocities, we observe a power-law behaviour of the coherence length, reminiscent of the Kibble-Zurek mechanism. However, we find exponents that strongly depend on the final interaction strength and thus lie outside the scope of this mechanism.
\end{abstract}

\maketitle


Phase transitions are ubiquitous but rather intricate phenomena and it took until the 20th century until a theory of classical phase transitions was established. Quantum phase transitions (QPTs) are marked by sudden drastic changes in the nature of the ground state upon varying a parameter of the Hamiltonian. They constitute one of the most intriguing frontiers of modern quantum many-body and condensed-matter physics \cite{sachdevbook,Transitions_in_focus_2008,Ions,tzker_Plenio_Zurek_et_al__2013,1311.1543,ZurekDornerZoller}. While it is typically possible to adiabatically follow the slowly changing ground state in a gapped phase, these spectral gaps usually close at a QPT. Since adiabaticity is therefore bound to break down, several important questions emerge: How does a state dynamically evolve across the QPT, i.e.\ how does the transition literally happen? To what extent can the static ground state of a gapless phase be prepared in a realistic finite-time experiment? When entering a critical phase associated with an infinite correlation length -- such as superfluid or ferromagnetic order, at what rate and by what mechanism will these correlations build up? Despite the fundamental importance of these questions, fully satisfactory answers have not been identified so far. While the intrinsic complexity of the underlying non-integrable models hinders numerical studies in most cases, the progress in the field of ultracold atoms now enables quantitative experiments in clean, well isolated, and highly controllable systems.

Here, we study for the first time the quantitative dynamics of a transition into a quantum critical phase in the regime of short and intermediate quench times. As a prototypical many-body system with a QPT we use the transition from a Mott insulator to a superfluid in the Bose-Hubbard model \cite{Fisher,Jaksch,Greiner,Zwerger} by changing (quenching) a parameter of the Hamiltonian. The non-equilibrium settings considered here are relatively well understood for sudden quenches, where the buildup of superfluid correlations (coherence) can be described by the ballistic spreading of quasiparticles \cite{Cheneau,Cheneau2012}. These excitations are generated during the instantaneous parameter change and spread with a group velocity limited by Lieb-Robinson bounds \cite{LiebRobinson}. For continuous quenches the situation is substantially more complex, since the continuous change of the Hamiltonian leads to drastically different elementary excitations throughout the evolution. Moreover, a continuous quench typically starts in the ground state, and the relevant excitations are only created during the ramp. While there is a large body of literature trying to capture these intricate dynamics of creation and change of quasiparticles in terms of scaling laws \cite{DziarmagaReview,sondhi_scaling}, a comprehensive and fully satisfactory theory is lacking and many questions are still largely open. These descriptions are built on, e.g., adiabatic perturbation theory \cite{PolkovnikovReview,polkovnikov_universal} or scaling collapses \cite{sondhi_scaling,Kolodrubetz,1311.1543}. Free models allow an exact treatment \cite{cherng_levitov,deng_viola} and can help to build an intuition for more complex physical systems. The Kibble-Zurek framework \cite{Kibble,Zurek,ZurekDornerZoller,DziarmagaReview,1310.1600} provides a simple guideline for the growth of correlations and predicts the density of defects following asymptotically slow ramps. It is, however, still not satisfactorily understood which correlation length results from crossing a phase transition in a strongly correlated model at a finite rate. The situation is aggravated by the fact that in higher-dimensional lattice systems, the available numerical techniques do not allow an accurate classical numerical simulation of this setting for long evolution times. 

\begin{figure}[htb]
	\centering
		\includegraphics[width=84mm]{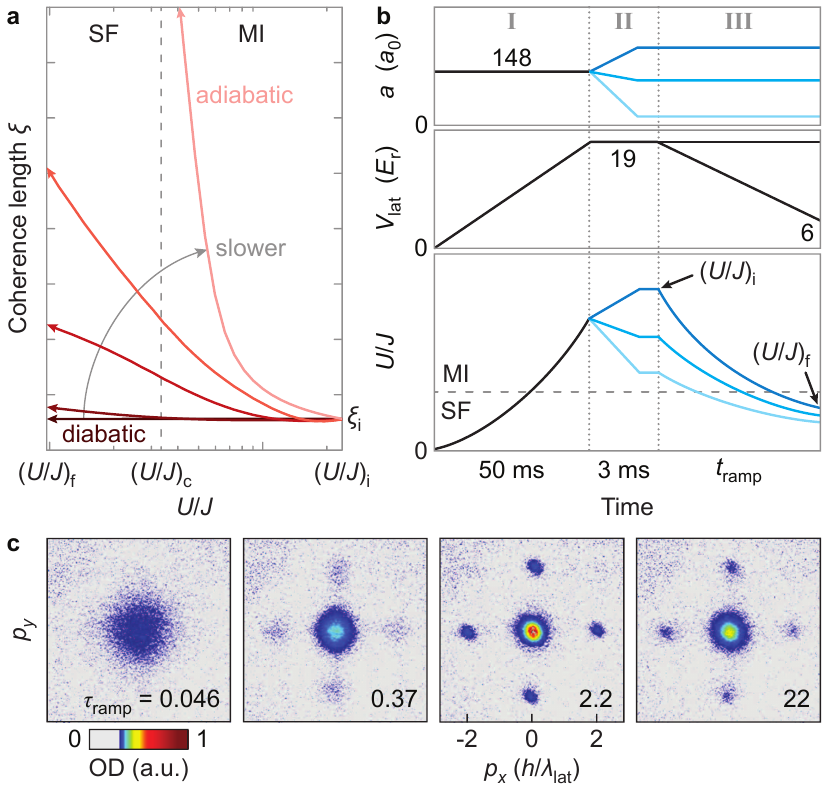}
    \caption{\textbf{Growth of coherence length during a quench and experimental sequence.} \textbf{a}, Numerical calculation of the evolution of the coherence length $\xi$ when ramping from a strong interaction $(U/J)_{\text{i}}$ in the Mott insulating regime (MI, right) over the critical point $(U/J)_{\text{c}}$ to $(U/J)_{\text{f}}$ in the superfluid regime (SF, left) in a homogeneous 1D system. The colors indicate the quench velocity, from fast (dark red) to the infinitely slow adiabatic limit (light red). \textbf{b}, Experimental sequence for scattering length $a$ (in units of the Bohr radius $a_0$), lattice depth $V_{\text{lat}}$, and $U/J$. During (\textbf{I}) we prepare a large central Mott insulator with unity filling. The different scattering length values $a$ chosen in (\textbf{II}) lead to different initial $(U/J)_{\text{i}}$ and final $(U/J)_{\text{f}}$ values for the final lattice ramp in (\textbf{III}), performed in variable time $t_{\text{ramp}}$. In 1D (2D), only one (two) lattice directions are reduced in the final lattice ramp. The horizontal dashed line indicates the critical $(U/J)_{\text{c}}$, separating the superfluid from the Mott insulating regime. \textbf{c}, Recorded time-of-flight absorption images for $(U/J)_{\text{f}}=3.2$ ($(U/J)_{\text{i}}=110$) in 2D for several $\tau_{\text{ramp}}$ (main text).}
	\label{eoc_sequence}
\end{figure}

In this work, we use ultracold atoms in an optical lattice to study the Mott to superfluid transition in the Bose-Hubbard model for experimental timescales far away from the adiabatic limit. We extract the coherence length from the width of the interference peaks in time-of-flight absorption images and observe that, as expected, the final coherence length depends strongly on the quench rate (Fig.\ \ref{eoc_sequence}a): While the resulting coherence length should diverge in the limit of adiabatic ramps, fast quenches result in short coherence lengths. We are able to probe this phase transition experimentally in \mbox{one-,} two-, and three-dimensional systems (1D, 2D, 3D), as well as for negative absolute temperature states \cite{Braun}. We compare our measurements in the 1D case with a numerical analysis and find excellent agreement.

Our experiments (Fig.\ \ref{eoc_sequence}b and Methods for details) started by (I) loading a large $n=1$ Mott insulator of $^{39}\text{K}$ atoms in a 3D optical lattice of depth $V_{\text{lat}}=V_{\text{i}}=19E_{\text{r}}$ at $U/J\geq 250$ close to the atomic limit of having a product state with exactly one atom per site. Here, $E_{\text{r}}=h^2/(2m\lambda_{\text{lat}}^2)$ denotes the recoil energy with Planck's constant $h$, the atomic mass $m$, and the lattice wavelength $\lambda_{\text{lat}}=736.65\,\text{nm}$. The on-site interaction energy of the Bose-Hubbard Hamiltonian \cite{Braun} is denoted by $U$ and the tunnelling matrix element by $J$. In the deep lattice (II), the scattering length was then tuned within a wide range of values via a Feshbach resonance at a magnetic field of $B=402.50G$ \cite{Zaccanti}, resulting in different values of the initial interaction strength $(U/J)_{\text{i}}$ in the deep lattice. We have verified numerically that this Feshbach ramp is very close to adiabatic such that, within the central Mott insulator, the state at this point can be assumed to be the ground state of the system (Supplementary Section G). Following this state preparation, the Mott to superfluid phase transition was crossed (III) by linearly ramping down the lattice depth along the horizontal $x$-direction to $V_{\text{lat}}^x=V_{\text{f}}=6E_{\text{r}}$ in variable times $t_{\text{ramp}}$ ($V_{\text{lat}}(t)=V_{\text{i}}+(V_{\text{f}}-V_{\text{i}})\cdot t/t_{\text{ramp}}$), resulting in a smaller interaction strength $(U/J)_{\text{f}}$ in the final shallow lattice. For experiments in 2D and 3D, we simultaneously ramped down the lattice depth along both horizontal directions or all three directions, respectively.

\begin{figure}[htb]
	\centering
		\includegraphics[width=84mm]{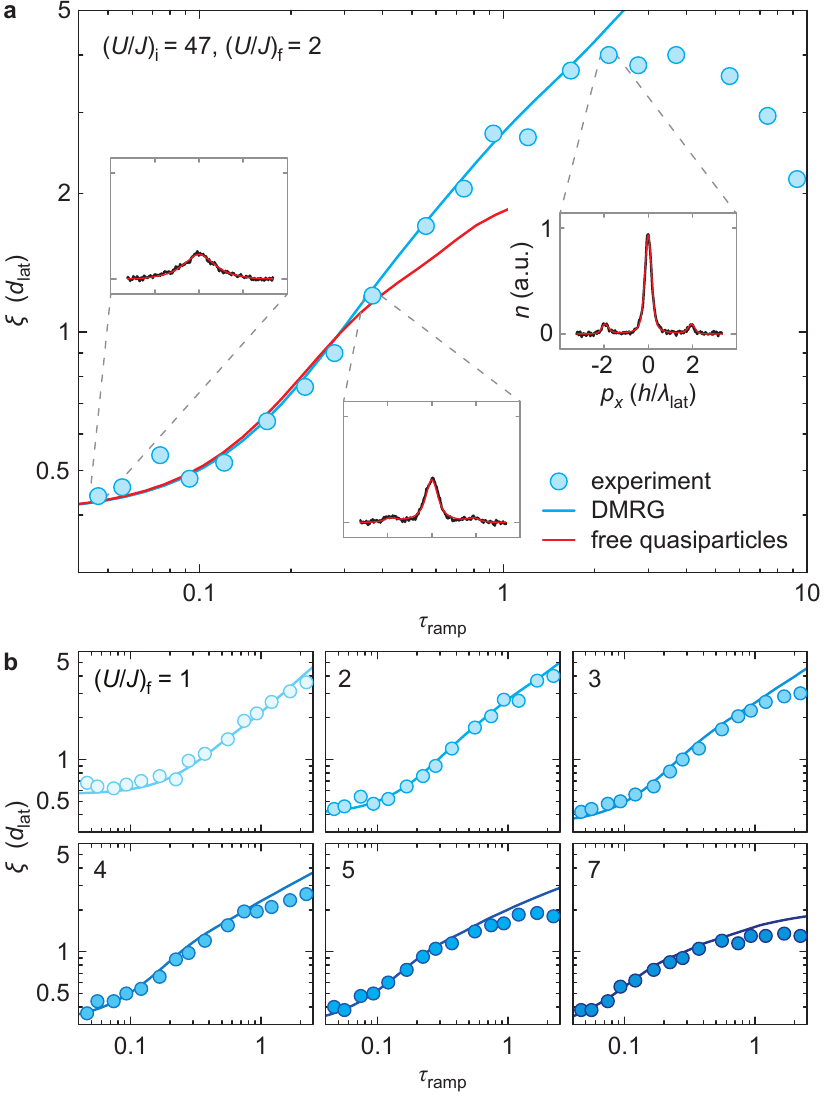}
    \caption{\textbf{Emergence of coherence for increasing ramp times.} \textbf{a}, Insets show integrated absorption images (black) for the 1D sequence with corresponding fits of a modelled interference pattern (red, Methods). The resulting coherence lengths $\xi$ are shown in the main figure versus $\tau_{\text{ramp}}$ in a double-logarithmic plot. The blue curve shows a DMRG calculation and the red curve the doublon-holon free quasiparticle model (Supplementary Section F) \textbf{b}, $\xi$ versus $\tau_{\text{ramp}}$ for several $(U/J)_{\text{f}}$ values in 1D. Points indicate experimental data, solid curves DMRG data. Throughout this work, the ramp time is measured in integrated tunneling times (main text).}
	\label{xi_and_power_law}
\end{figure}

After the ramp, we immediately switched off all trapping potentials and recorded absorption images along the vertical $z$-direction after a time-of-flight of $t_{\text{TOF}}=7\text{ms}$ (Fig.\ \ref{eoc_sequence}c). From the width of the interference peaks, we extracted the coherence length of the system, i.e.\ the characteristic length scale of an exponential decay of correlations, by calculating the expected time-of-flight profiles for various coherence lengths and fitting them to the experimental data (Fig.\ \ref{xi_and_power_law}a and Methods). We measure the number of tunneling times during the ramp by defining a dimensionless ramp time $\tau_{\text{ramp}}=t_{\text{ramp}}\cdot 2\pi\bar{J}/h\approx t_{\text{ramp}}\cdot 0.93/\text{ms}$. Here, $\bar{J}=\int_{V_{\text{i}}}^{V_{\text{f}}}J(V)\,\text{d}V/(V_{\text{f}}-V_{\text{i}})$ denotes the average tunneling rate during the ramp. We focus on the short and intermediate ramp time regime, where mass transport is negligible and the dynamics is governed by the behaviour of the homogeneous system at the multi-critical tip of the Mott lobe \cite{Fisher}. This experiment captures for the first time the physics of essentially homogeneous quantum systems entering a critical phase. In contrast, previous work \cite{DeMarco} investigated the generic transition through the side of the Mott lobe, which is typical for inhomogeneous systems and is dominated by mass transport, studied the inverse superfluid to Mott insulator transition \cite{bakr_greiner}, the vacuum to superfluid transition \cite{Zhang} or the transition of spinor Bose-Einstein condensates to a ferromagnetic state \cite{Sadler}.

The experimentally measured coherence length (Fig.\ \ref{xi_and_power_law}) displays several distinct dynamical regimes. For very fast ramps, the evolution can be approximated as being sudden, and the measured coherence length $\xi$ essentially equals that of the initial Mott insulator $\xi_{\text{i}}$. The latter is significantly below one lattice spacing $d_{\text{lat}}=\lambda_{\text{lat}}/2$ and increases for smaller $(U/J)_{\text{i}}$ closer to the critical point at $(U/J)_{\text{c}}\approx 3.3$ in 1D \cite{Carrasquilla_Rigol}. For larger $\tau_{\text{ramp}}$, $\xi$ quickly increases up to several lattice spacings. For $\tau_{\text{ramp}}\gtrsim 2-5$, the fitted $\xi$ starts to decrease again due to the influence of the trap: Contrary to a homogeneous system, the equilibrium distributions of both density and entropy density in a trapped system depend strongly on the interaction strength. While strong interactions result in a large Mott insulating core with constant density, surrounded by a superfluid or thermal shell at lower density, a weakly interacting superfluid is described by a parabolic Thomas-Fermi distribution. Intuitively speaking, the density distribution cannot equilibrate during fast and intermediate lattice ramps and results in gradients in the chemical potential, which give rise to dephasing between lattice sites that increases over time and becomes relevant for slower ramps, $\tau_{\text{ramp}}\gtrsim 2-5$ (Supplementary Section D). Furthermore, while entropy in an ideal bosonic Mott insulator is located predominantly in the surrounding non-insulating shell, it is distributed more homogeneously in a superfluid. Thus, for short times the trapped system is indistinguishable from the homogeneous system, while for longer times trapping effects dominate the dynamics \cite{Bernier_Kollath,Natu_Hazzard_Mueller}.

The measured emergence of coherence observed here is indeed a generic feature of the \textit{homogeneous} Bose-Hubbard model, i.e.\ without trap, as can be seen by directly comparing the experimental data to a classical density matrix renormalization group (DMRG) \cite{Schollwock201196} simulation of the homogeneous system, based on matrix-product states (Fig.\ \ref{xi_and_power_law}b) \cite{opentebd}. The sole input parameters for the simulation were $U$ and $J$ and no fitting to the experimental data points was performed. By extensive scaling in bond dimension as well as Trotter step size, we have ensured numerical convergence. Further cross-validation was obtained by an optimised exact diagonalisation code performing a Runge-Kutta numerical integration of a homogeneous Bose-Hubbard model on 15 sites (Supplementary Section E). We find excellent agreement between experiment and numerical data for small and intermediate ramp times up to $\tau_{\text{ramp}}\approx 1$. For larger $\tau_{\text{ramp}}$, the coherence length of the simulated homogeneous system continues to increase, while the experimental data starts to decrease due to the trap.

In the fast and slow limits, the physics of the continuous quench in the homogeneous model can be understood from two complementary viewpoints: For fast ramps, $\tau_{\text{ramp}}\lesssim 0.2$, the dynamics can be well described in terms of ballistically spreading quasiparticles, implemented in a doublon-holon fermionic model (Fig.\ \ref{xi_and_power_law}a and Supplementary Section F). In this picture, fermionic excitations are continuously created during the quench and spread with their corresponding velocity that can be shown to follow a Lieb-Robinson bound \cite{LiebRobinson,CalabreseCardy06,CramerEisert,EisertOsborne06,BravyiHastingsVerstraete06,CramerEisertScholl08,Kollath08}. For intermediate ramps, $\tau_{\text{ramp}}\gtrsim 0.2 $, interactions between the quasiparticles become important. For slow ramps, on the other hand, one can employ the language of adiabatic quenches and the adiabatic theorem: In the beginning of the ramps, the gap is sufficiently large and we can expect the system to perfectly follow the change of the ground state. Closer to the phase transition, this approximation breaks down and the full complexity of the problem emerges. If the breakdown of the adiabatic approximation occurs sufficiently close to the critical point, where the physics is governed by scaling laws, the Kibble-Zurek mechanism suggests a power-law growth of the correlation length, where the exponent is governed by the critical exponents of the quantum phase transition \cite{Zurek,1310.1600}.

\begin{figure}[htb]
	\centering
		\includegraphics[width=84mm]{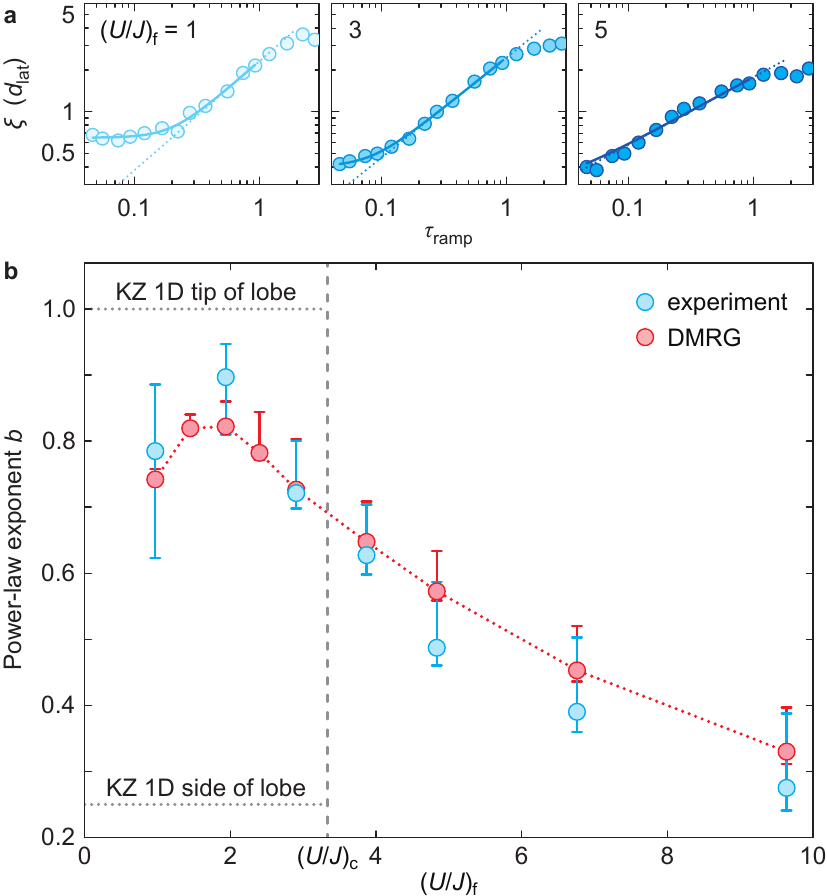}
    \caption{\textbf{Power-law increase of coherence length.} \textbf{a}, Power-law fits to experimental data in 1D. The fit function $\xi(\tau_{\text{ramp}})=(\xi_i^4+(a\,\tau_{\text{ramp}}^b)^4)^{1/4}$ heuristically includes the initial coherence $\xi_i$ (Supplementary Section C) and is applied to all ramp times up to $\tau_{\text{ramp}}^{\text{max}}=1.0$. The dotted line shows the pure power-law $\xi(\tau_{\text{ramp}})=a\,\tau_{\text{ramp}}^b$ for the above fitted parameters. \textbf{b}, Exponents $b$ for experiment (blue) and DMRG (red), extracted via identical fitting procedures. The error bars include the effect of various fitting ranges as well as the fitting uncertainties (Supplementary Section C). The red dotted line guides the eye. The vertical dashed line indicates the critical value $(U/J)_{\text{c}}\approx 3.3$ for the Mott-superfluid transition and the horizontal dotted lines indicate the predictions $b=1$ and $b=1/4$ \cite{Batrouni} of a typical Kibble-Zurek (KZ) model at the tip or side of the Mott lobe, respectively (main text and Supplementary Section G).}
	\label{experiment_exact_diagonalisation}
\end{figure}

We find that, also in our setting, the evolution can be captured in terms of simple power-laws (Fig.\ \ref{experiment_exact_diagonalisation}): Within a range of $\tau_{\text{ramp}}$ of around one order of magnitude, the growth of the coherence length is very well approximated by a power-law,
\begin{equation}
	\xi(\tau_{\text{ramp}})=a\,\tau_{\text{ramp}}^b.
\end{equation}
This is rather surprising, as rough estimates suggest that the above adiabaticity condition is not fulfilled in the intermediate ramp time regime studied in this work. By using fits to the experimental or numerical data, we extract the exponents $b$ (Fig.\ \ref{experiment_exact_diagonalisation}a,b), finding values that are always substantially lower than $b=1$ suggested by earlier theoretical works based on the Kibble-Zurek mechanism for the transition at the tip of the $n=1$ Mott lobe \cite{cucchietti_bh}. More refined studies of this 1D transition, which is of the Kosterlitz-Thouless type, show that for realistic experimental scales, smaller power-laws are expected \cite{1312.5139}. Our main finding on the dynamics of the 1D phase transition, however, cannot be captured by a simple scaling model, as the observed exponent crucially depends on the final point $(U/J)_{\text{f}}$ of the quench. In the experiment, ramps with different final values $(U/J)_{\text{f}}$ also have different initial interactions $(U/J)_{\text{i}}$. Since the first part of the evolution is, however, essentially adiabatic, this change of the initial interaction does not significantly alter the emerging scaling laws (Supplementary Section G). Due to the rather small resulting coherence lengths, we can also rule out finite-size effects as the origin for this behaviour, as further corroborated by numerical simulations on systems of various sizes (Supplementary Section E). The simulations also show that the trap cannot be the reason for the $(U/J)_{\text{f}}$-dependence of the exponents, since the homogeneous model considered in the numerics and the experimental data agree extremely well and the influence of the trap is only visible for ramp times $\tau_{\text{ramp}}\gtrsim 1$ (Supplementary Section D). An inhomogeneous Kibble-Zurek scaling has recently been analysed for a classical phase transition in ion chains \cite{Ions,tzker_Plenio_Zurek_et_al__2013} and for quantum \cite{DeMarco} as well as thermal \cite{Scherer,Lamporesi} phase transitions in ultracold atom systems. In contrast, the agreement between the inhomogeneous experiment and the numerics for the homogeneous system shown here proves that we effectively probe the multi-critical quantum phase transition of the homogeneous Bose-Hubbard model, not influenced by trap effects.

\begin{figure}[htb]
	\centering
		\includegraphics[width=84mm]{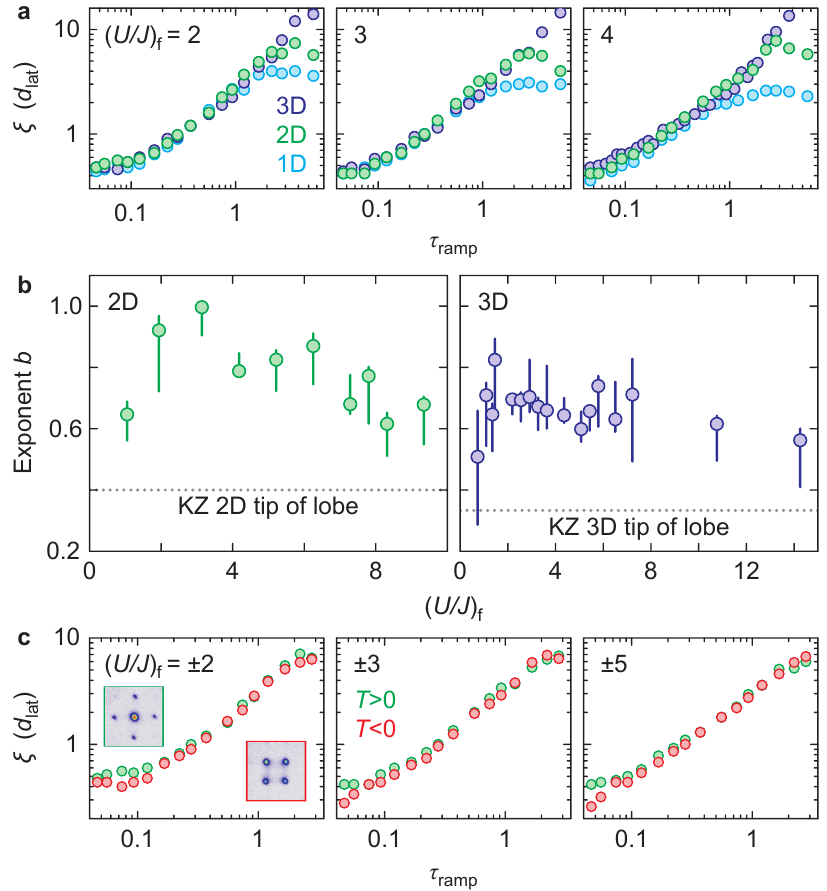}
    \caption{\textbf{Emergence of coherence in higher dimensions and for negative absolute temperature.} \textbf{a}, Experimental data for 1D, 2D, and 3D for various $(U/J)_{\text{f}}$. \textbf{b}, Exponents for the 2D and 3D case, extracted from power-law fits to the experimental data with the identical fitting procedure as in Fig.\ \ref{experiment_exact_diagonalisation}. The dotted lines indicate the Kibble-Zurek predictions $b=0.4$ and $b=1/3$ for the tip of the Mott lobes in the 2D and 3D case, respectively. \textbf{c}, Experimental data for the 2D case for positive and negative absolute temperature for various $(U/J)_{\text{f}}$. The insets in the left panel show TOF images for both cases at $\tau_{\text{ramp}}=2.2$.}
	\label{pos_T_neg_T_and_dimensions}
\end{figure}

Fig.\ \ref{pos_T_neg_T_and_dimensions} shows the results of corresponding experiments for the 2D and 3D Bose-Hubbard model, which are inaccessible to analytical models as well as current numerical tools. After having verified that the observed quantum dynamics in 1D indeed agree with the homogeneous Bose-Hubbard model, the experiments in 2D and 3D can be regarded as analogue quantum simulations in a regime out of reach of classical simulation using known methods. Interestingly, the data for higher dimensions show similar power-laws as the 1D case, even though any critical scaling analysis would strongly depend on dimensionality. Thus, we again find that the dynamics of the Mott to superfluid phase transition shows complex behaviour on the studied intermediate timescale, that simple approaches based on the critical exponents alone, such as KZM, cannot fully capture. While the extracted exponents for the most part increase for decreasing interaction strength, they start to decrease again for $(U/J)_{\text{f}}\lesssim 2$ in all dimensions (Fig.\ \ref{pos_T_neg_T_and_dimensions}b). Furthermore, the full coherence dynamics for $\tau_{\text{ramp}}\lesssim 1$ appears to be almost independent of dimensionality and is mainly governed by the final interaction $(U/J)_{\text{f}}$. Therefore, in the regime where $\xi$ has increased only up to a few $d_{\text{lat}}$, the influence of dimensionality on the spreading of correlations is marginal. Higher-dimensional systems continue the power-law behaviour for longer ramp times than in lower dimensions. This deviation might be explained by the different critical values $(U/J)_{\text{c}}$ in 1D, 2D and 3D: For the same final $(U/J)_{\text{f}}$ value, the quench in the 1D system ends closer to, or even deeper in, the Mott regime than for higher dimensions, limiting the maximum achievable coherence length in addition to the dephasing effect in the trap. A rigorous comparison between different dimensions would have to involve a detailed analysis of the ramp schemes as well as the different mass flow predictions and the individual critical values together with the different scaling of the equilibrium correlations, i.e.\ quasi-long-range order in 1D versus true long-range order in higher dimensions.

To show that the timescale for the emergence of coherence is not influenced by any possible remnant phase order in the initial state that might seed the dynamics, we additionally studied the emergence of coherence in the attractive Bose-Hubbard model: By crossing the Feshbach resonance and additionally inverting the external confinement in the deep lattice ((II) in Fig.\ \ref{eoc_sequence}b), we can realize an attractive Mott insulator at negative absolute temperature \cite{Braun,Mosk,Rosch,Nagerl} (Supplementary Section G). In Fig.\ \ref{pos_T_neg_T_and_dimensions}c, we compare the emergence of coherence between attractive and repulsive interactions in 2D, and find essentially identical behaviour. Deviations become visible only for strong interactions (not shown) and can most likely be attributed to multi-band effects \cite{luhmann_self-trapping_2008,Best}. Since positive and negative temperature superfluids occupy completely different quasimomenta with different correlations (Fig.\ \ref{pos_T_neg_T_and_dimensions}c, insets), we conclude that the emergence of coherence observed in the experiment is truly governed by the generic behaviour of the continuous quench. 

In conclusion, by performing an experimental quantum simulation we have studied the emergence of coherence across a QPT for various interactions, dimensionalities, and positive and negative absolute temperatures. In 1D, we have also performed a detailed theoretical and numerical analysis, and found very good agreement between experiment and DMRG calculations. The observed dynamics goes beyond the regime of free quasiparticles and, despite its complexity, we find that a simple scaling law emerges in a regime where neither the adiabatic theorem nor Lieb-Robinson bounds can characterize the evolution. While this power-law is reminiscent of a Kibble-Zurek type scaling, we find that, in the studied intermediate ramp time regime, the exponent crucially depends on $(U/J)_{\text{f}}$ and depends much less on dimensionality than suggested by the Kibble-Zurek mechanism.

This work raises the question how well the dynamical features of a quantum phase transition in complex models can generally be captured in terms of simple scaling laws by systematically expanding either the free quasiparticle picture or starting from the Kibble-Zurek mechanism: The success of the latter for slow quenches in a variety of specific models suggests that, compared to a full solution of the model, much less knowledge may be sufficient to characterise the evolution. A satisfactory answer to this question will be crucial for a theory of the dynamics of quantum phase transitions. Since exact numerical techniques are not available in higher dimensions, this work may inspire a deeper and more systematic analysis of the computational power of analogue quantum simulators in general. For example, it seems timely to identify, in the language of complexity theory, the precise way in which quantum simulators are indeed more powerful, even in the absence of error correction, than their classical analogues, and how accurately experimental quantum simulators can ultimately be certified as functioning quantum devices.

\vspace{\vspacelarge}

\section*{Methods}
The experiments started with essentially pure condensates of, depending on data set, $(25-85)\cdot 10^3$ bosonic $^{39}$K atoms in an oblate dipole trap with trap frequencies of $\omega=2\pi\cdot(50,50,181)\text{Hz}$ along the $(x,y,z)$ direction. We linearly ramped up a 3D optical lattice to a depth of $V_{\text{lat}}=19E_{\text{r}}$ (I in Fig.\ \ref{eoc_sequence}b). In the 1D case, we then quickly increased the transverse lattice depth to $V_{\text{lat}}^y=V_{\text{lat}}^z=30E_{\text{r}}$ to minimise correlations along the $y$- and $z$-directions. The scattering length during this loading procedure was $a=148a_0$, resulting in $(U/J)_x\approx 350$ (1D) or $U/J\approx 270$ (2D, 3D), i.e.\ deep in the Mott insulating regime close to the atomic limit. he trap frequency was increased during the loading to $(94,94,159)\text{Hz}$ in the 1D and 2D case and $(78,78,227)\text{Hz}$ in the 3D case to ensure a large Mott insulating region in the centre of the cloud.

The momentum distribution of the atoms in the optical lattice, typically probed using absorption imaging after long time-of-flight (TOF), is given by \cite{Gerbier,PhysRevA.78.013627}
\begin{equation}
	\label{formula_interference_pattern}
	\left\langle\hat{n}(\mathbf{k})\right\rangle=\frac{1}{{\cal N}}|\tilde{w}(\mathbf{k})|^2S(\mathbf{k}),
\end{equation}
with the Fourier transform of the on-site Wannier function $\tilde{w}(\mathbf{k})$ determining the overall envelope of the interference pattern and a normalisation factor ${\cal N}$. The interference term $S(\mathbf{k})$ has the form of a discrete Fourier transform and is given by a sum over all lattice sites at positions $\mathbf{r}_{\mu}$ and $\mathbf{r}_{\nu}$,
\begin{equation}
	S(\mathbf{k})=\sum_{\mathbf{r}_\mu,\mathbf{r}_\nu}e^{i\mathbf{k}(\mathbf{r}_\mu-\mathbf{r}_\nu)}\langle\hat{a}^{\dagger}_{\mu}\hat{a}_{\nu}\rangle,
\end{equation}
where $\hat{a}^{\dagger}_{\mu}$ and $\hat{a}_{\mu}$ are the creation and annihilation operators, respectively, for a boson on site $\mu$.

In the experiment, we probe the momentum distribution using a \textit{finite} time-of-flight $t_{\text{TOF}}=7\text{ms}$, and attribute a momentum $\mathbf{k}=m\mathbf{r}_{\text{TOF}}/(\hbar t_{\text{TOF}})$ to each position $\mathbf{r}_{\text{TOF}}$ in real space. Due to the finite time-of-flight, the initial density distribution still influences this measured distribution and the interference term is generalized to \cite{Gerbier}
\begin{equation}
  \label{TOF}
	\tilde{S}(\mathbf{k})=\sum_{\mathbf{r}_\mu,\mathbf{r}_\nu}e^{i\mathbf{k}(\mathbf{r}_\mu-\mathbf{r}_\nu)-i\frac{m}{2\hbar\,t_{\text{TOF}}}(\mathbf{r}_\mu^2-\mathbf{r}_\nu^2)}\langle\hat{a}^{\dagger}_{\mu}\hat{a}_{\nu}\rangle.
\end{equation}
The second term in the exponential provides a correction to a pure Fourier transform and is equivalent to the quadratic term in the Fresnel approximation of near-field optics. We model the correlators by assuming a Gaussian in situ density distribution with width $R$ and exponentially decaying correlations between lattice sites:
\begin{equation}
   \begin{aligned}
\label{TOF2}
\langle\hat{a}^{\dagger}_{\mu}\hat{a}_{\nu}\rangle_{(T>0)}&=&\sqrt{n_{\mu}}\sqrt{n_{\nu}}\cdot\exp{\left(-\frac{|\mathbf{r}_\mu-\mathbf{r}_\nu|}{\xi}\right)}\\
&=&\exp\left(-\frac{\mathbf{r}_\mu^2+\mathbf{r}_\nu^2}{4R^2}-\frac{|\mathbf{r}_\mu-\mathbf{r}_\nu|}{\xi}\right)
   \end{aligned}
\end{equation}
Here, $\xi$ denotes the coherence length and $n_{\mu}$ the density at site $\mu$. In the case of negative temperatures, the correlator contains an additional phase term,
\begin{equation}
\label{negT}
\langle\hat{a}^{\dagger}_{\mu}\hat{a}_{\nu}\rangle_{(T<0)}=\langle\hat{a}^{\dagger}_{\mu}\hat{a}_{\nu}\rangle_{(T>0)}\cdot e^{i\boldsymbol{\pi}(\mathbf{r}_{\mu}-\mathbf{r}_{\nu})}.
\end{equation}

To extract the coherence length in the system, we integrate the TOF images over a small region of width $d_{\text{int}}\approx 0.2ht_{\text{TOF}}/\lambda_{\text{lat}}m$ along the $y$-direction. We fit the resulting interference pattern with calculated patterns of the above model for various $\xi$ and fixed $R$ and extract the coherence length $\xi$ from the fit (Supplementary Section A). We determine $R$ independently by fitting a Gaussian distribution to in situ images (Supplementary Section B). Sample fits are shown as insets of Fig.\ \ref{xi_and_power_law}a for the 1D case. Even though this rather simple ansatz cannot reproduce the numerically calculated correlation functions in detail (Supplementary Section E), it is sufficient to reproduce the experimentally measured interference patterns. Extracted coherence lengths are shown in the main plot of Fig.\ \ref{xi_and_power_law}a. In the case of the 2D and 3D sequences, correlations also spread in the transverse directions. To extract the coherence length, we integrate the images in the same range of diameter $d_{\text{int}}$ along the $y$-direction as for 1D and fit the calculated 1D interference patterns to the resulting data.

\section*{Acknowledgments}
\vspace{\vspacesmall}
\noindent
We thank Marc Cheneau, Masud Haque, Corinna Kollath, Michael Kolodrubetz, Alessandro Silva, Christian Go\-go\-lin, Lode Pollet, and Wojciech H.\ Zurek for helpful discussions, Salvatore Manmana for helpful insights into numerical aspects and Tim Rom for assistance in setting up the experiment. We acknowledge financial support by the Deutsche Forschungsgemeinschaft (FOR801, Deutsch-Israelisches Kooperationsprojekt Quantum phases of ultracold atoms in optical lattices), the US Defense Advanced Research Projects Agency (Optical Lattice Emulator program), Nanosystems Initiative Munich, the Spanish MINECO project FIS2011-29287, the Spanish JAE predoc program, CAM research consortium QUITEMAD S2009-ESP-1594, the EU (SIQS, RAQUEL), the ERC (TAQ), FQXi, Studienstiftung des Deutschen Volkes, and the BMBF (QuOReP).

%
%

\renewcommand{\theequation}{S\arabic{equation}}
\renewcommand{\thefigure}{S\arabic{figure}}
\setcounter{equation}{0}
\setcounter{figure}{0}

\vspace{0.5cm}


\appendix

\noindent
\section*{Supplementary Information}

\section{A: Ramp details and extraction of coherence length}
The employed final lattice ramp can be well approximated by an exponential function
\begin{equation}
(U/J)(t)=B\cdot a\cdot e^{-C(t/t_{\text{ramp}})^D},
\end{equation}
where $a$ is the scattering length and $t\in [0,t_{\text{ramp}}]$. The parameters depend on dimensionality and are obtained by fits to the real ramps. In the 1D case, they are $B=2.33/a_0$, $C=3.04$, $D=1.10$, in 2D, $B=1.80/a_0$, $C=3.32$, $D=1.11$, and in 3D, $B=1.79/a_0$, $C=3.60$, $D=1.11$.

\begin{figure}[htb]
	\centering
		\includegraphics[width=84mm]{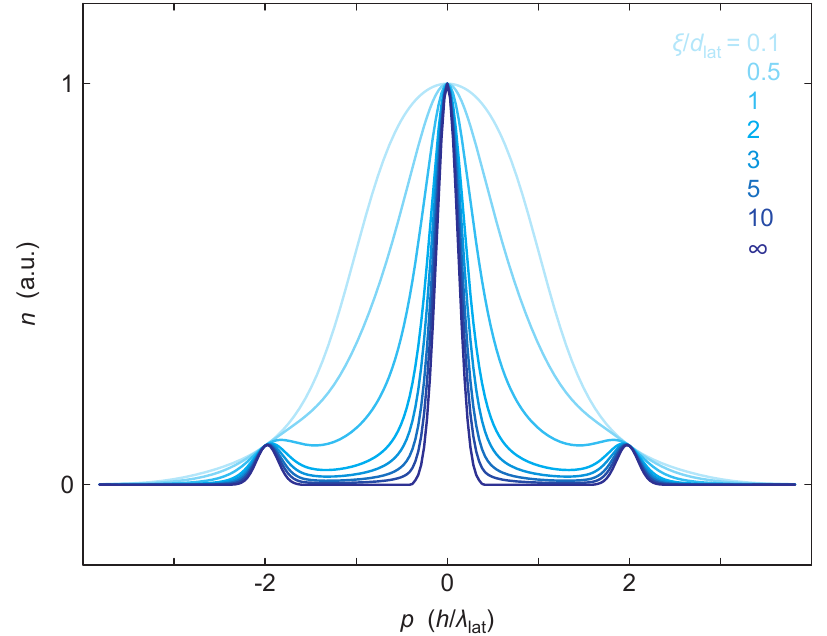}
    \caption{Calculated normalized density interference patterns following Eqs.\ (\ref{formula_interference_pattern}-\ref{TOF2}) in the main text for $R=31d_{\text{lat}}$ and various coherence lengths $\xi$.}
	\label{fig:fig_S_theory_curves_R_31}
\end{figure}

Fig.\ \ref{fig:fig_S_theory_curves_R_31} shows several calculated one-dimensional interference patterns (Eqs.\ (\ref{formula_interference_pattern},\ref{TOF},\ref{TOF2})\,) for a fixed width $R$ and for the experimentally relevant time-of-flight $t_{\text{TOF}}=7\text{ms}$. For very small coherence lengths, $\xi\ll d_{\text{lat}}$, only the Fourier transform of the on-site Wannier function as envelope function is visible. For larger coherence lengths, the interference pattern becomes more and more pronounced. The width of the interference peaks, however, saturates at a minimum value which is given by the in situ width $R$ of the cloud.

\begin{figure*}[htb]
	\centering
		\includegraphics[width=\textwidth]{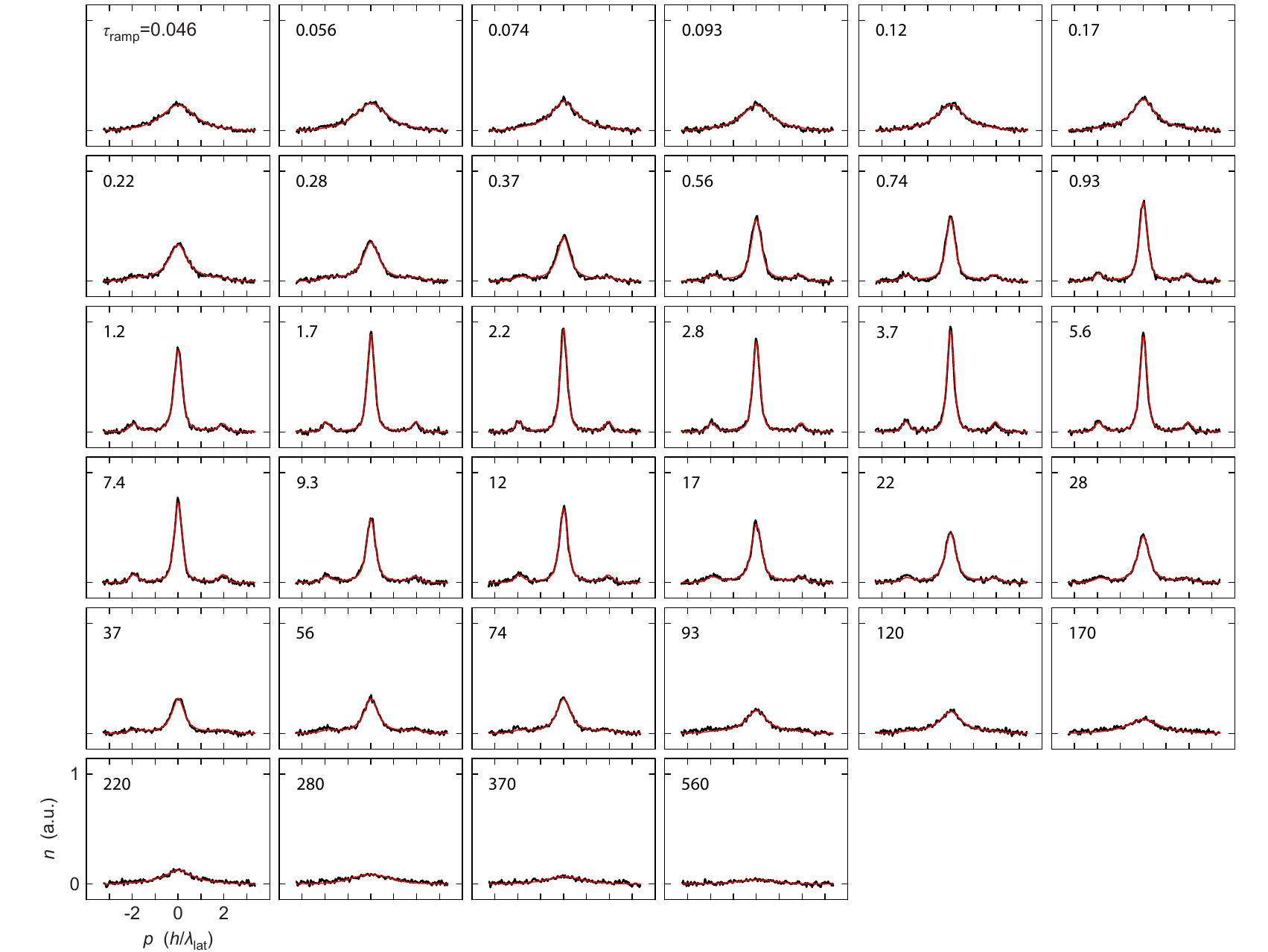}
    \caption{Comparison of experimental data integrated over a small region along $y$ (black, see Methods), with the fitted calculated interference pattern (red) for $(U/J)_{\text{f}}=2$ in 1D at positive temperature, for various ramp times. For this plot, after fitting the normalized curves, the curves have been rescaled to the original amplitudes of the experimental curves.}
	\label{fig:fig_S_xi_fits}
\end{figure*}

To extract the coherence length $\xi$ of a measured interference pattern, we normalised the interference pattern and determined the sum of absolute values of residuals (SAR) between the measured and each calculated interference pattern and determined the coherence length $\xi$ by finding the minimal SAR. Fig.\ \ref{fig:fig_S_xi_fits} shows that, for all ramp times, the rather simple model of Eqs.\ (\ref{formula_interference_pattern},\ref{TOF},\ref{TOF2}) can reproduce the experimentally measured interference curves well. We applied the fitting method also to an almost pure condensate at a scattering length of $a=37a_0$ that was loaded into a 3D lattice of depth $V_{\text{lat}}=6E_{\text{r}}$. The resulting SAR values indicate a lower bound of the coherence length of $\xi\gtrsim 15d_{\text{lat}}$ but are compatible with an infinite coherence length, as expected for a mostly superfluid state. In contrast to phase retrieval algorithms \cite{Kosior}, where arbitrary phase profiles $\phi(\mathbf{r})$ can be reconstructed from TOF images in the case of a single pure wavefunction $\psi(\mathbf{r})=|\psi(\mathbf{r})|\exp(i\phi(\mathbf{r}))$ with finite support, we fix the phase of the correlators but do not require the system to be in a pure state.

\section{B: Extraction of cloud radius}
\label{radius}
\begin{figure}[htb]
	\centering
		\includegraphics[width=84mm]{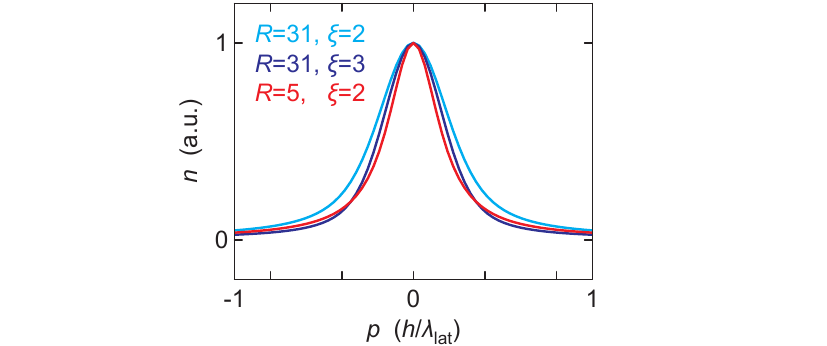}
	\caption{Calculated interference pattern for various widths $R$ and coherence lengths $\xi$. Both an increasing $\xi$ and a decreasing $R$ lead to a decreasing peak width in the interference pattern.}
	\label{fig:fig_S_theory_curves_dependence_of_xi_and_R}
\end{figure}

The fitting function for the extraction of the coherence lengths (Eqs.\ (\ref{formula_interference_pattern}-\ref{negT})\,) contains two free parameters, the coherence length $\xi$ and the in situ cloud width $R$. Their influence on the fit is not completely independent as both a larger $\xi$ and a smaller $R$ lead to narrower peaks in the resulting interference pattern (Fig.\ \ref{fig:fig_S_theory_curves_dependence_of_xi_and_R}). We therefore did not extract both parameters simultaneously from the fit, but instead determined $R$ independently from in situ images or from the total atom number.

\begin{figure}[htb]
	\centering
		\includegraphics[width=84mm]{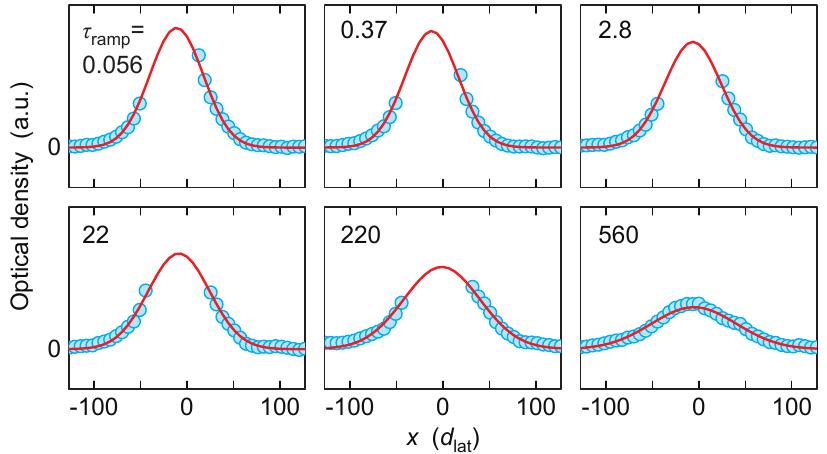}
	\caption{Fitting of in situ clouds. The blue points are cuts through experimentally measured in situ distributions for $(U/J)_{\text{f}}=2$ in 1D at positive temperature for various ramp times. In some images, the high density of the cloud centre lead to strong absorption of the imaging beam such that the signal cannot be distinguished from noise. We excluded the corresponding data points from the fits and also removed them from the plot. The red curves are cuts through the fitted two-dimensional Gaussian functions.}
	\label{fig:fig_S_insitu_fits}
\end{figure}

For selected $(U/J)_{\text{f}}$ values in 1D, 2D (each at positive and negative temperature) and 3D, we recorded in situ images of the atomic clouds for all ramp times in (III) in Fig.\ 1b, from which we directly determined $R$. In situ imaging was done by freezing out the atomic distribution via a rapid increase of the 3D optical lattice to $V_{\text{lat}}=33E_{\text{r}}$. During a hold time of several ms, we switched off the magnetic field and subsequently performed absorption imaging of the frozen atomic distribution. Fig.\ \ref{fig:fig_S_insitu_fits} shows several cuts through in situ images, together with Gaussian fits
\begin{equation}
n(x,y)=Ae^{-\frac{(x-x_0)^2}{(2R_x^2)}-\frac{(y-y_0)^2}{(2R_y^2)}}
\end{equation}
to the full images, from which we determine the root mean square width $R=(R_x^2+R_y^2)^{1/2}$.

\begin{figure}[htb]
	\centering
		\includegraphics[width=84mm]{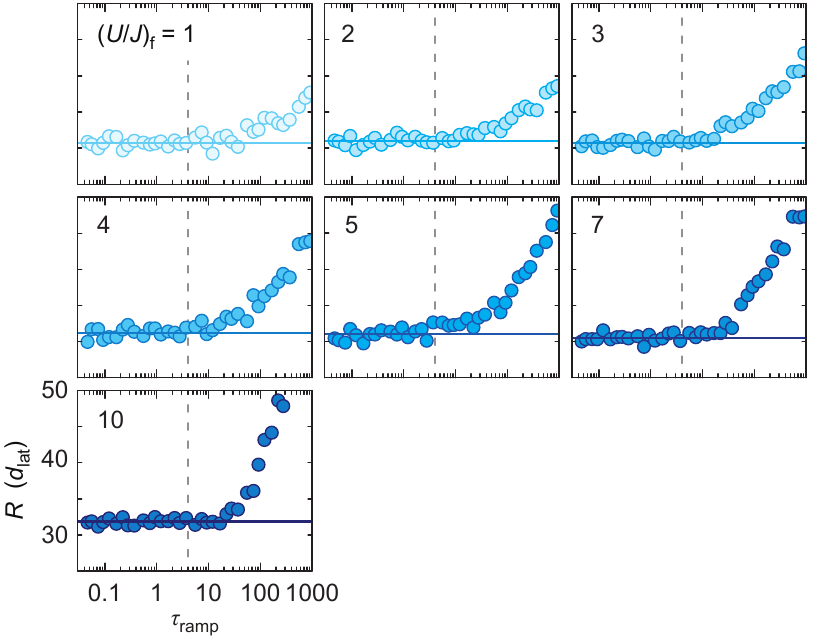}
	\caption{Fitted in situ widths $R=(R_x^2+R_y^2)^{1/2}$ versus ramp time for various $(U/J)_{\text{f}}$ values in 1D at positive temperature. The vertical dashed line at $\tau_{\text{ramp}}=4$ indicates the maximum ramp time up to which the fitted $R$ were averaged. The resulting average is indicated by the solid horizontal line.}
	\label{fig:fig_S_R_vs_tramp}
\end{figure}

The resulting widths are plotted in Fig.\ \ref{fig:fig_S_R_vs_tramp} as a function of ramp time for all $(U/J)_{\text{f}}$ values in the 1D measurement. For small to intermediate $\tau_{\text{ramp}}$, the width of the in situ cloud is constant. Only for slower ramps, $\tau_{\text{ramp}}\gtrsim 10$, global mass redistribution becomes relevant and the cloud expands during the lattice ramp. As we are mainly interested in the short and intermediate timescale dynamics, we fix the width for each $(U/J)_{\text{f}}$ value by averaging all $R$ for $\tau_{\text{ramp}}\leq 4$ and rounding the result to an integer number of lattice constants.

\begin{figure}[htb]
	\centering
		\includegraphics[width=84mm]{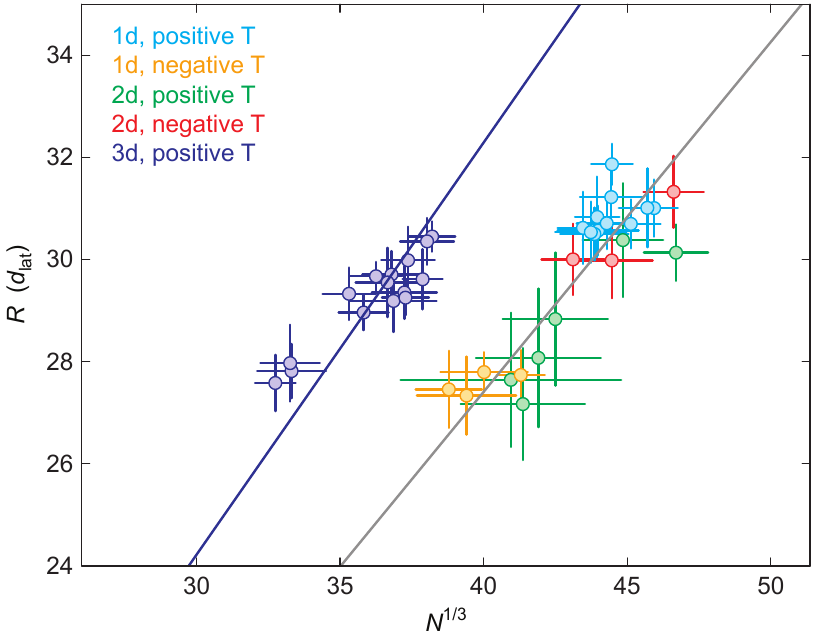}
	\caption{Fitted in situ widths $R$ versus $N^{1/3}$. The linear fit $R=m\cdot N^{1/3}$ to the combined 1D and 2D data (grey line) yields a slope of $m_{\text{1D,2D}}=0.685(4)d_{\text{lat}}$ and the fit to the 3D data (purple line) $m_{\text{3D}}=0.807(5)d_{\text{lat}}$. The different scalings in the two cases are caused by different trap frequencies.}
	\label{fig:fig_S_R_vs_N}
\end{figure}

For those $(U/J)_{\text{f}}$ values for which we did not record in situ images, we determined $R$ indirectly from the total atom number $N$, which we measured in TOF images via simple area sums. As one might expect, for those $(U/J)_{\text{f}}$ values for which we recorded both TOF and in situ images, the width $R$ is strongly correlated with $N^{1/3}$ (Fig.\ \ref{fig:fig_S_R_vs_N}). In the experiment, we used identical preparation schemes for all 1D and 2D runs, but different trap frequencies in the 3D case, and therefore obtain different dependencies for these two sets. By fitting the function $R=m\cdot N^{1/3}$ to all available in situ data, we can relate the measured atom number $N$ to the width $R$ also for those $(U/J)_{\text{f}}$ values for which we only recorded TOF images. We round each resulting $R$ value to an integer number of lattice constants. All resulting $R$ presented in this work lie between $26$ and $32d_{\text{lat}}$.

\section{C: Extracting power-law exponent}
To obtain the most precise value of the exponent of the power-law growth of the coherence length, it is desirable to include as many data points as possible in the fit. When fitting a pure power-law $\xi(\tau_{\text{ramp}})=a\,\tau_{\text{ramp}}^b$ to the data, however, the fitting range is limited by two effects: For very small ramp times, the coherence length is dominated by the initial coherence length, $\xi\approx\xi_{\text{i}}$, and for large ramp times, $\tau_{\text{ramp}}\gtrsim 1$, the influence of the trap leads to a deviation from the power-law behaviour. To reduce the problem of determining the `correct' fitting range, we use a heuristic fitting procedure that includes the initial coherence length $\xi_{\text{i}}$ and fit all data points up to a maximum ramp time $\tau_{\text{ramp}}^{\text{max}}$. We choose $\tau_{\text{ramp}}^{\text{max}}=1.0$ as it guarantees that, for all data sets available, no data points are included that are strongly influenced by the trap (Supplementary Section D).

\begin{figure}[htb]
	\centering
		\includegraphics[width=84mm]{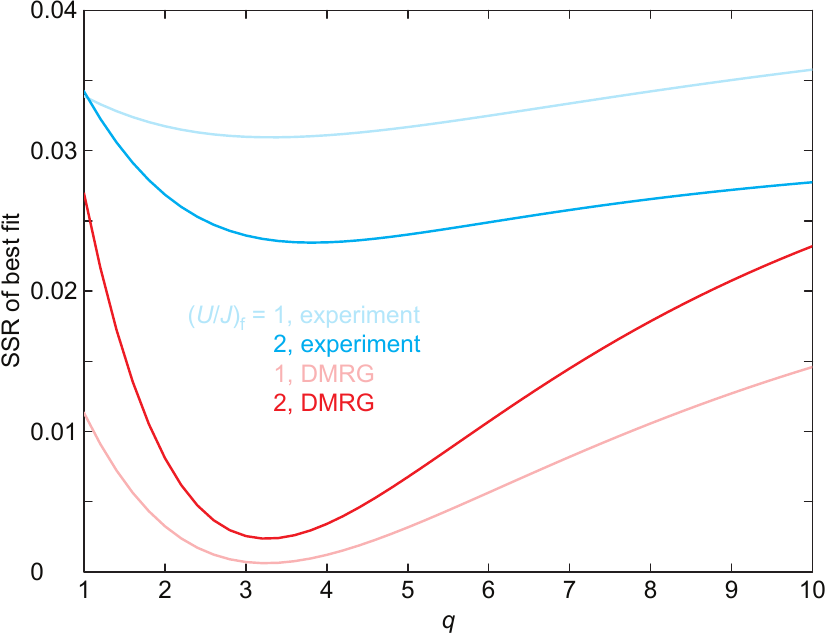}
  \caption{SSR of the general power-law fit versus parameter $q$ for 1D data. Light-blue and dark-blue lines indicate the result for fits to experimental data for $(U/J)_{\text{f}}=1$ and $2$, respectively. Light-red and dark-red lines are the analogue results for DMRG data. A choice of $q=4$ leads to a close-to-minimum SSR for all four cases.}
	\label{fig:fig_S_optimizing_q}
\end{figure}

\begin{figure}[htb]
	\centering
		\includegraphics[width=84mm]{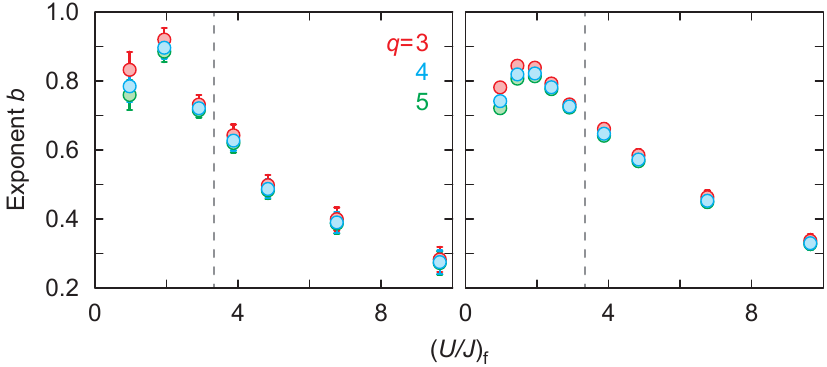}
	\caption{Fitted exponents in 1D for $\tau_{\text{ramp}}^{\text{max}}=1$ for various values of $q$. Left, experimental data, right, DMRG data. Error bars are fit uncertainties and the vertical dashed line indicates $(U/J)_{\text{c}}$. For details about the fitting procedure, see Figs.\ \ref{fig:fig_S_1d_fits_below_UJc} and \ref{fig:fig_S_1d_fits_above_UJc}.}
	\label{fig:fig_S_exponents_various_q}
\end{figure}

\begin{figure}[htb]
	\centering
		\includegraphics[width=84mm]{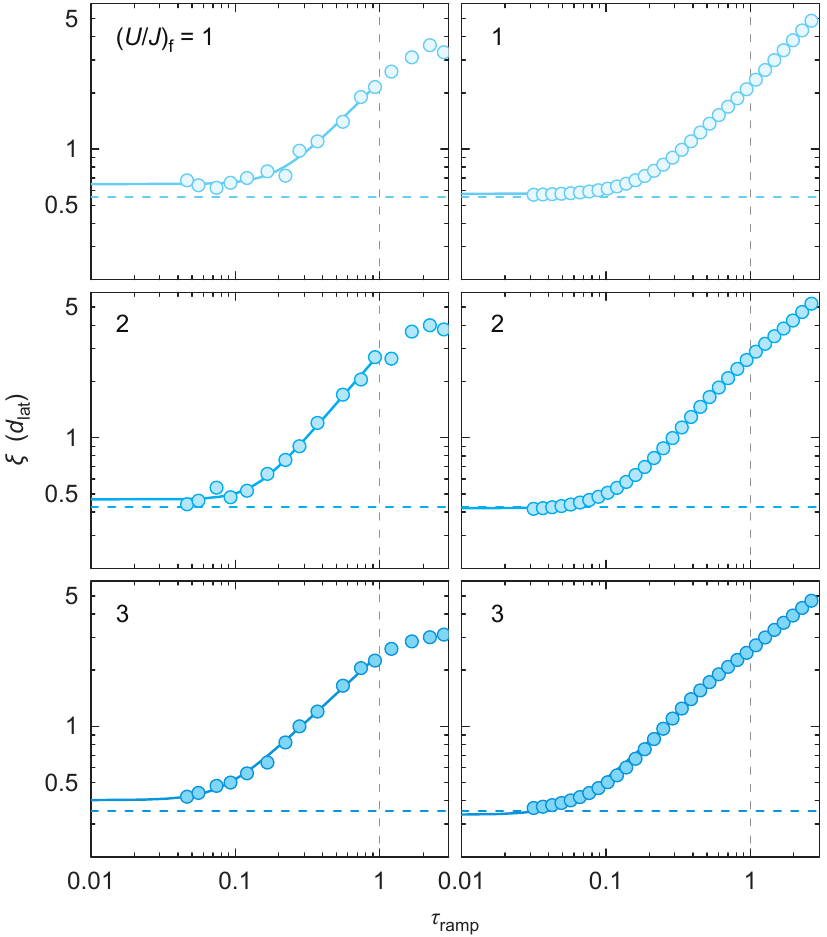}
	\caption{Power-law fits ($q=4$) including the initial coherence length $\xi_{\text{i}}$ as a free parameter in 1D for values $(U/J)_{\text{f}}<(U/J)_{\text{c}}$. Left, experimental data, right, DMRG data. The vertical dashed line indicates the upper end of the fitting range, $\tau_{\text{ramp}}^{\text{max}}=1.0$. The horizontal dashed lines indicate the numerically calculated $\xi_{\text{i}}$ at the beginning of the lattice ramp, showing good agreement with the extrapolated fitted values $\xi(\tau_{\text{ramp}}\rightarrow 0)$.}
	\label{fig:fig_S_1d_fits_below_UJc}
\end{figure}

The applied heuristic function
\begin{equation}
\xi(\tau_{\text{ramp}})=\left(\xi_{\text{i}}^q+(a\,\tau_{\text{ramp}}^b)^q\right)^{1/q}
\end{equation}
smoothly interpolates between $\xi_{\text{i}}$ and the power-law growth. Within the power-law regime, the coherence created during the ramp is substantially larger than $\xi_{\text{i}}$. Thus, for $q\geq 2$, the initial coherence has little influence on the power-law exponent and this fitting procedure yields the same power-law exponent as choosing the fitted range by eye. The precise value of the parameter $q$ is fixed by fitting both numerical and experimental data in the 1D case for the values $(U/J)_{\text{f}}=2$ and $1$ with various $q$. To match the data as closely as possible, we aim to minimize the sum of squared residuals (SSR) of the fit; Fig.\ \ref{fig:fig_S_optimizing_q} indicates that $q=4$ is a good compromise. Fig.\ \ref{fig:fig_S_exponents_various_q} shows that the extracted exponents are robust with respect to the particular choice of $q$ and Fig.\ \ref{fig:fig_S_1d_fits_below_UJc} illustrates that a choice of $q=4$ indeed captures the emergence of coherence very well.

\begin{figure}[htb]
	\centering
		\includegraphics[width=84mm]{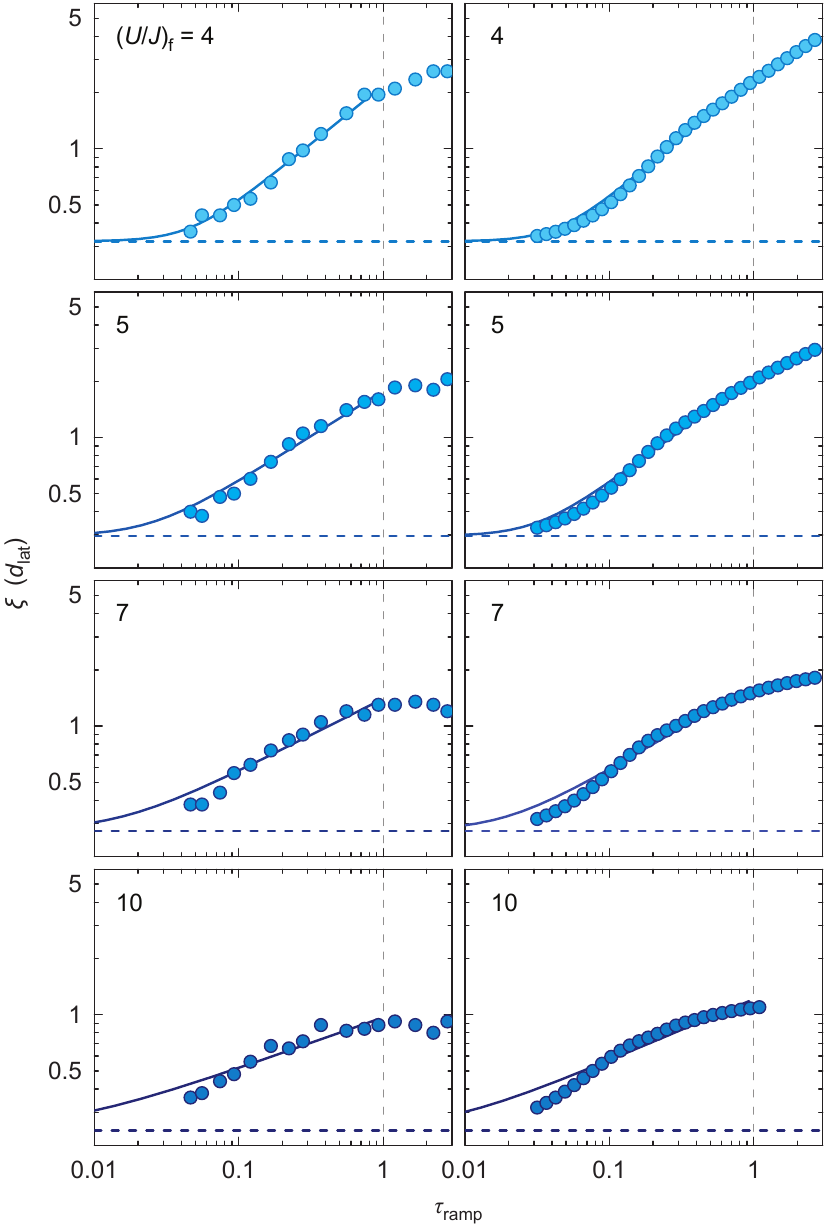}
	\caption{Power-law fits for $q=4$ in 1D for values $(U/J)_{\text{f}}>(U/J)_{\text{c}}$. Left, experimental data, right, DMRG data. The vertical dashed line indicates the upper end of the fitting range, $\tau_{\text{ramp}}^{\text{max}}=1.0$. The initial coherence length $\xi_{\text{i}}$ is fixed to the respective calculated value, indicated by the horizontal dashed line.}
	\label{fig:fig_S_1d_fits_above_UJc}
\end{figure}

In general, the power-law exponent as well as $\xi_{\text{i}}$ are extracted directly from fitting the data. This method is robust since the influence of the two parameters on the fit function is independent. Only for very strong interactions in 1D where the system remains in the Mott insulating regime, the power-law increase is rather slow and it is difficult to distinguish the power-law regime from the regime that is dominated by $\xi_{\text{i}}$. To optimize the stability of the fit for those particular data sets where $(U/J)_{\text{f}}>(U/J)_{\text{c}}$, we fix $\xi_{\text{i}}$ to the numerically calculated values. Fig.\ \ref{fig:fig_S_1d_fits_below_UJc} shows that the calculated $\xi_{\text{i}}$ of the state at the beginning of the lattice ramp is close to the fitted $\xi_{\text{i}}$ in the experimental data where $(U/J)_{\text{f}}<(U/J)_{\text{c}}$, indicating that fixing the initial coherence length has little influence on the extracted power-law exponent. The resulting power-law fits capture the data well (Fig.\ \ref{fig:fig_S_1d_fits_above_UJc}); only for very large $(U/J)_{\text{f}}\gg(U/J)_{\text{c}}$, systematic deviations from the simple power-law model become relevant.

\begin{figure}[htb]
	\centering
		\includegraphics[width=84mm]{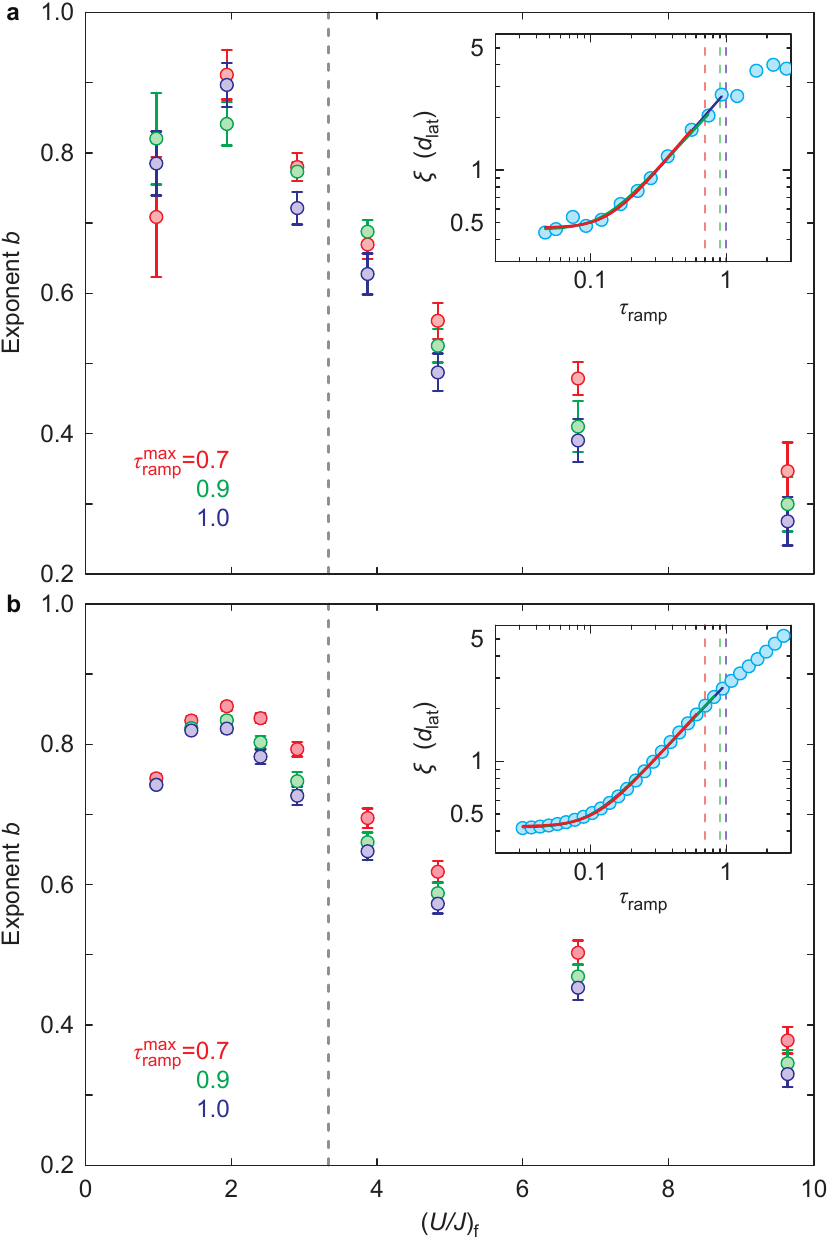}
	\caption{Fitted exponents in 1D for various limits $\tau_{\text{ramp}}^{\text{max}}$ of the fit range for \textbf{a}, experimental data, and \textbf{b}, DMRG data.	Error bars are fit uncertainties and the vertical dashed line indicates $(U/J)_{\text{c}}$. The insets show sample fits for $(U/J)_{\text{f}}=2$, with the various $\tau_{\text{ramp}}^{\text{max}}$ indicated by the vertical dashed lines.}
	\label{fig:fig_S_exponents_all_fit_ranges}
\end{figure}

While this method eliminates the problem of setting the lower limit of the fitting range, the upper limit $\tau_{\text{ramp}}^{\text{max}}$ is still arbitrary. To quantify the dependence of the fitted power-law exponent on the choice of $\tau_{\text{ramp}}^{\text{max}}$, we also performed power-law fits for different $\tau_{\text{ramp}}^{\text{max}}$. A larger value than $\tau_{\text{ramp}}^{\text{max}}=1.0$ would lead to systematic errors due to the dephasing effects of the trap. Therefore, we perform the power-law fit to all data sets for reduced limits of $\tau_{\text{ramp}}^{\text{max}}=0.9$ and $0.7$. In Fig.\ \ref{fig:fig_S_exponents_all_fit_ranges}, all resulting exponents in the 1D case are plotted, including fit errors. To represent the uncertainty with respect to the choice of $\tau_{\text{ramp}}^{\text{max}}$, we take the total amplitude of fitting errors for all three different $\tau_{\text{ramp}}^{\text{max}}$ as error bar, i.e.\ the range from the maximum point of the highest error bar to the minimum point of the lowest error bar (see Fig.\ 3 in the main text).

\section{D: Influence of the trap}
\label{trap}
In contrast to the homogeneous model system, the experiment takes place in a harmonic trap. As outlined in the main text (Fig.\ \ref{xi_and_power_law}b), the emergence of coherence shows a discrepancy between the homogeneous theory and the experiment for relatively long ramp times, $\tau_{\text{ramp}}\gtrsim 1$.

\begin{figure}[htb]
  \centering
  \includegraphics[width=84mm]{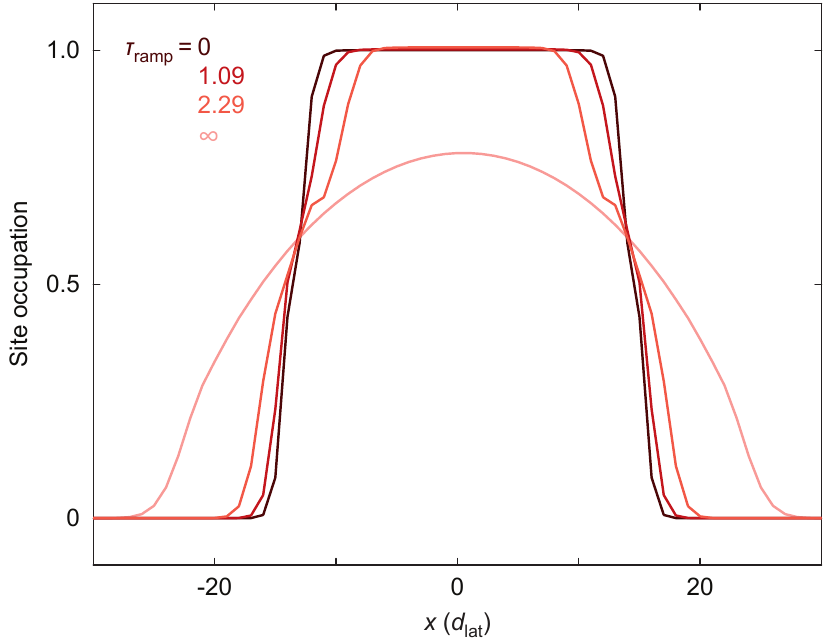}
  \caption[quench schedule]{Density profile for various ramp times and the ground state both at the beginning (diabatic limit $\tau_{\text{ramp}}=0$) and at the end (adiabatic limit $\tau_{\text{ramp}}=\infty$) of the ramp. The simulation was performed for 28 particles and a ramp with $(U/J)_{\text{f}}=3$ in 1D for a trap frequency of $\omega_x/2\pi=66\text{Hz}$.}
  \label{fig:trap_density}
\end{figure}

To understand the influence of the trap, let us first consider the equilibrium situation of the Mott insulating and superfluid phases for systems with and without trap. In the homogeneous case, both the Mott insulator as well as the resulting superfluid state share the same filling, e.g.\ one atom per lattice site, $n=1$. Thus, when performing quenches between the two regimes, no mass or entropy redistribution is required. In the trapped system, in contrast, the density and entropy distributions strongly depend on the phase of the ensemble. While a weakly interacting superfluid is described by a parabolic Thomas-Fermi distribution, an $n=1$ Mott insulator has a flat central density, with a superfluid or thermal shell at lower density around the Mott insulating core. While entropy in a strongly interacting ($U/J\gg (U/J)_{\text{c}}$) bosonic Mott insulator state is, for low temperatures, located only in the surrounding shell, it is distributed more homogeneously in the weakly interacting system. Therefore, during a quench across the phase transition in a trapped system, not only phase coherence has to be established or destroyed, but also mass and entropy has to be redistributed, as illustrated in Fig.\ \ref{fig:trap_density}.

In our experiment, we use a 50ms ramp to initially load the atoms into a deep optical lattice with depth $V_{\text{lat}}=19E_{\text{r}}$. This timescale, in contrast to shorter ramps, experimentally turned out to produce large Mott insulating cores with low doublon fraction \cite{Ronzheimer}, thus allowing the bulk of necessary mass and entropy transport. Since the ramp is, however, not perfectly adiabatic, the ensemble is heated and additional entropy is created during the ramp and will accumulate in the non-insulating outer shell during this slow loading procedure. The density distribution at the beginning of the lattice ramp therefore consists of a low-entropy $n=1$ Mott insulating core surrounded by a hot thermal gas at lower density. The Mott core may also carry some entropy in the form of holes and it cannot be assumed that the ensemble is in global thermal equilibrium at the end of the initial loading ramp. Nonetheless, the remarkable agreement between the measured coherence lengths and the numerical predictions (Fig.~\ref{xi_and_power_law} in the main text) clearly indicates that the dynamical behaviour of the experimental system at short and intermediate ramp times is dominated by that of a perfect Mott insulator.

During the final lattice ramp-down from $V_{\text{lat}}=19E_{\text{r}}$ to $6E_{\text{r}}$, substantial mass and entropy redistributions would be necessary for an adiabatic evolution. Fig.\ \ref{fig:fig_S_R_vs_tramp}, however, indicates that relevant mass transport happens only on a timescale of $\tau_{\text{ramp}}\gtrsim 10$. Since the initial loading corresponds to a ramp time of $\tau_{\text{ramp}}\approx 32$ for the used ramp, this behaviour is consistent with our choice of the loading ramp.

\begin{figure}[htb]
	\centering
		\includegraphics[width=84mm]{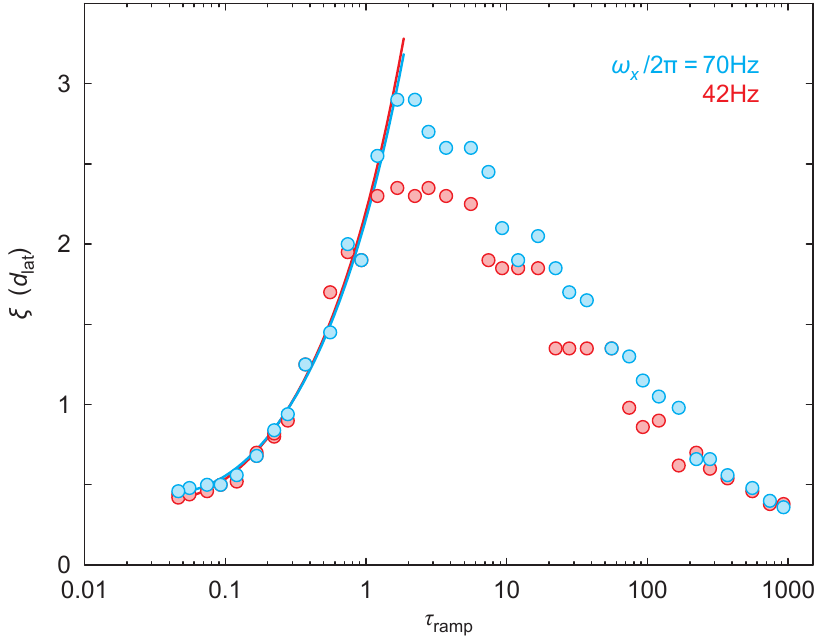}
	\caption{Coherence length versus ramp time for $(U/J)_{\text{f}}=3$ in 1D in a log-lin plot for two different trap frequencies $\omega_x$ along the $x$-axis. The solid curves are corresponding power-law fits up to $\tau_{\text{ramp}}^{\text{max}}=1.0$, plotted to larger $\tau_{\text{ramp}}$.}
	\label{fig:fig_S_eoc_various_traps}
\end{figure}

\begin{figure}[htb]
	\centering
		\includegraphics[width=84mm]{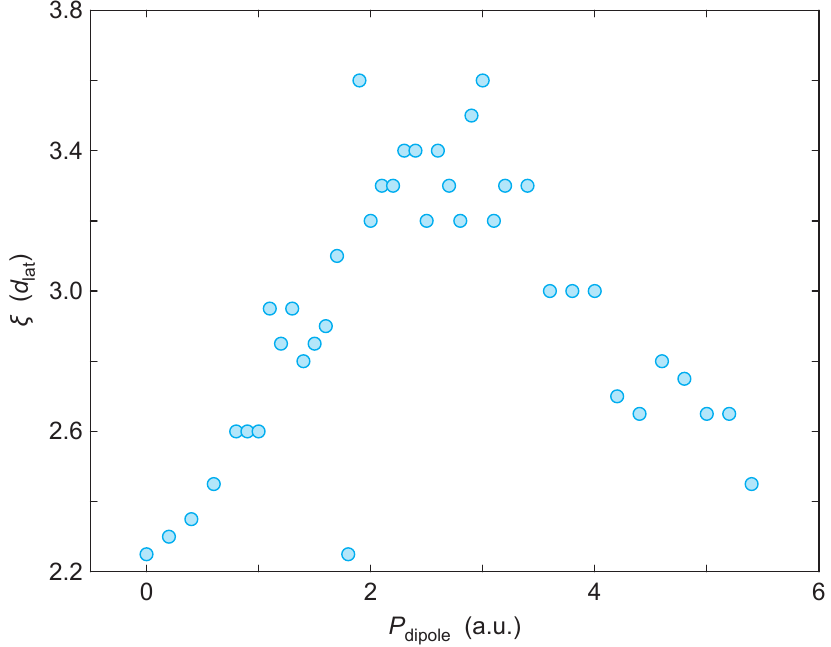}
	\caption{Coherence length for $(U/J)_{\text{f}}=4$ in 1D for a fixed ramp time $\tau_{\text{ramp}}=2.8$ versus power in the vertical dipole trap beam. The dipole power at the maximum corresponds to a trap frequency $\omega_x/(2\pi)\approx 70\text{Hz}$.}
	\label{fig:fig_S_xi_vs_z_dip_3ms}
\end{figure}

\begin{figure}[htb]
	\centering
  \includegraphics[width=84mm]{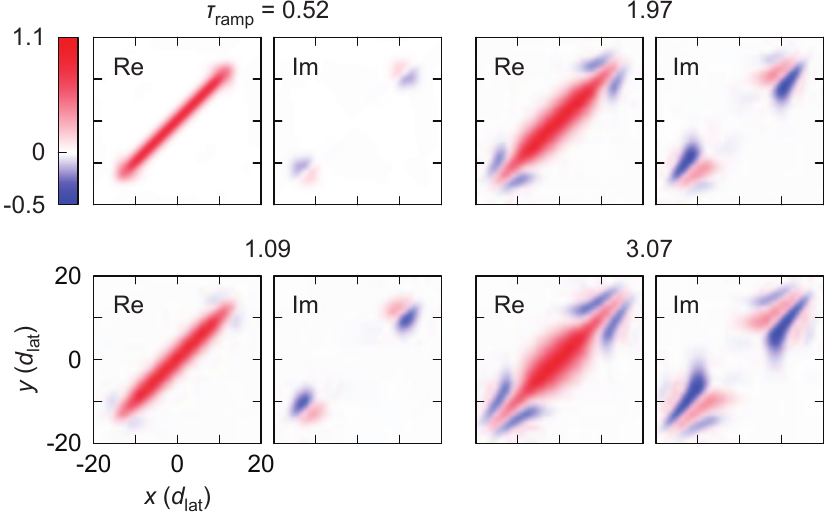}
  \caption{Final correlator for $(U/J)_{\text{f}}=3$ in trapped systems filled with $N=24$ particles for various ramp times. Real (Re) and imaginary part (Im) are plotted separately.}
	\label{fig:theory_trap_correlators}
\end{figure}

For short and intermediate lattice ramps, on the other hand, mass transport is negligible and the resulting final density distribution cannot correspond to the equilibrium density distribution of the superfluid state. The resulting chemical potential is thus not constant throughout the system. Intuitively, the decrease of the coherence length caused by the trap can be captured in a dephasing picture: The difference in chemical potential leads to a phase difference between lattice sites that increases linearly with time. These phase differences, which can be seen in the complex two-point correlators in Fig.\ \ref{fig:theory_trap_correlators}, in turn drive a particle current that tries to equilibrate the chemical potential and thereby establish an equilibrium density profile. This mechanism competes with the emergence of phase coherence as soon as the latter is being established. Since the phase difference accumulates over time, it is negligible for short times $\tau_{\text{ramp}}\lesssim 1$, where we can observe the power-law behaviour. For slower ramps, dephasing increases and leads to a deviation of the emergence of coherence signal from the power-law describing the homogeneous system. Fig.\ \ref{fig:fig_S_eoc_various_traps} shows the comparison of two measurements in 1D for the same $(U/J)_{\text{f}}$ but different trap frequencies. While the initial emergence of coherence is identical and gives rise to the same power-law exponents, the breakdown of the power-law occurs earlier for the weaker trap. This indicates that the breakdown of the power-law is indeed caused by the influence of the trap. Fig.\ \ref{fig:fig_S_xi_vs_z_dip_3ms} shows that the coherence length for a given $\tau_{\text{ramp}}$ can be optimised by choosing the right trapping frequency, therefore also maximising the range of the power-law regime: For some trapping frequency, for which the required mass redistribution is minimal, also the mismatch of the chemical potential across the system is minimised.

For a ramp time of $\tau_{\text{ramp}}\gtrsim 40$, the ramp should be close-to-adiabatic such that dephasing is not relevant. However, at this ramp time the coherence length is significantly reduced compared to the maximum value, by around a factor of 2. This can be attributed to entropy transport which is expected to happen on a similar timescale as mass transport: Only after this time, the entropy that is concentrated in the shell around the Mott insulating core has spread throughout the system. The increased entropy density reduces phase coherence between lattice sites. For very long ramp times $\tau_{\text{ramp}}\gtrsim 100$, we expect heating due to light scattering and technical noise as an additional effect that decreases the coherence length even further.

\begin{figure}[htb]
	\centering
  \includegraphics[width=84mm]{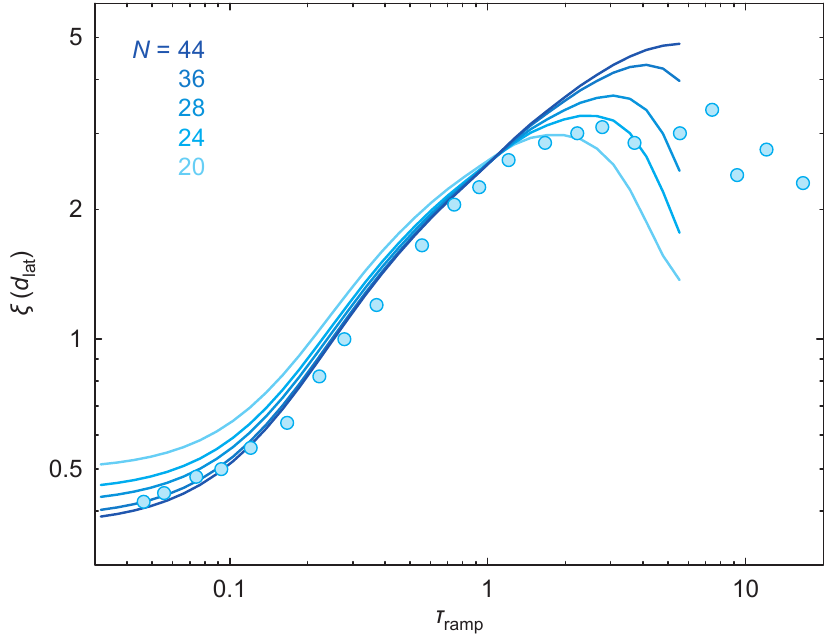}
  \caption{Emergence of coherence in the presence of a trap for $(U/J)_{\text{f}}=3$ in 1D. The points are experimental data and the solid curves DMRG calculations for various particle numbers $N$.}
	\label{fig:theory_trap}
\end{figure}

To qualitatively model the effect of the trap, we have performed DMRG simulations in the 1D case including a harmonic trap with a trap frequency of $\omega_x/2\pi=66\mathrm{Hz}$ (Supplementary Section E). The simulated 1D systems consist of tubes containing zero-temperature $n=1$ Mott insulators of variable lengths. The experimental data is an average over these tubes with different particle numbers, depending on the position within the ellipsoid created by the harmonic trap (see also Supplementary Material of Ref.\ \cite{Trotzky}). To extract the coherence length for each of the simulated tubes, we use the calculated density information and fit the two-point correlators in the same way as in Eq.\ \eqref{TOF2} in the main text. These calculations qualitatively confirm that the decrease of coherence length is indeed caused by the trap and that it is dominated by the influence of the shorter tubes (Fig.\ \ref{fig:theory_trap}). Any more elaborate modelling of the trapped situation is not feasible due to uncertainties in the initial state: Since the initial loading of the lattice is not perfectly adiabatic and the resulting initial state is not guaranteed to be in global thermal equilibrium, it is impossible to precisely predict the initial state without performing a full dynamical simulation of the complete 3D loading procedure, which is beyond current numerical techniques.

In Fig.\ \ref{fig:theory_trap_correlators}, we additionally plot the behaviour of the two-point correlators of a tube with few particles for different ramp times. For longer ramp times, $\tau_{\text{ramp}}>1$, the spatially dependent chemical potential gives rise to complex two-point correlators, illustrating the role of the trap.

\section{E: DMRG details and finite size effects}
\label{finite_size}
\begin{figure}[htb]
    \includegraphics[width=84mm]{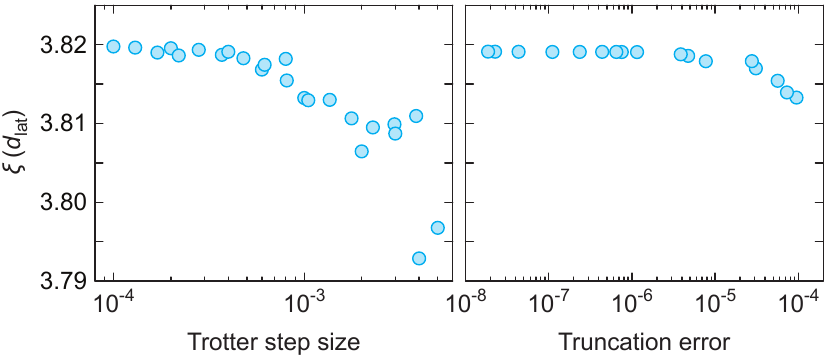}
    \caption[DMRG parameters]{Coherence length for $(U/J)_{\text{f}}=1$ in 1D for a fixed ramp time $\tau_{\text{ramp}}=2.0$ versus Trotter step size $\Delta_{\text{t}}$ (left) and truncation error (right), calculated for $N=40$. Based on this scaling, a Trotter step size of $\Delta_{\text{t}}=2\cdot10^{-4}$ was chosen. The bond dimension was taken sufficiently large to obtain a small enough truncation error $\epsilon_{\text{trunc}}<10^{-7}$.}
    \label{fig:DMRG}
\end{figure}

\begin{figure}[htb]
    \includegraphics[width=84mm]{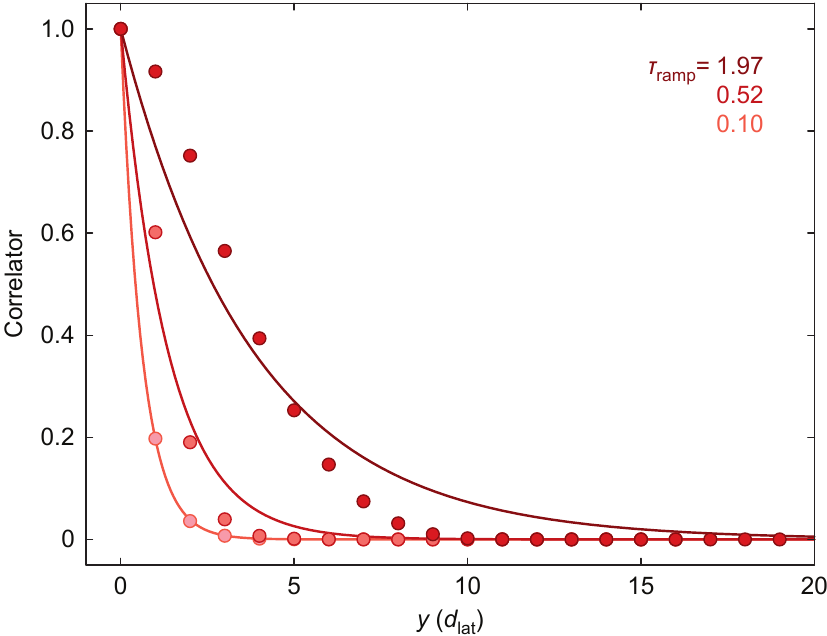}
    \caption[Exponential decay]{Spatial behaviour of the non-equilibrium correlators and the corresponding exponential fit for different ramp times for $(U/J)_{\text{f}}=1$ in 1D.}
    \label{fig:numerics_exponential_decay}
\end{figure}

The DMRG simulation is based on the `Open Source TEBD' code \cite{opentebd} that has been streamlined and optimized for this task. The only input parameters for the simulation are the time-dependent $U(t)$ and $J(t)$ and no fitting to the experimental data points was performed. The numerical simulations start in the ground state of the system. This turns out to be an excellent approximation for the experiment: The Feshbach ramp used to prepare the initial state is very close to adiabatic, since the preparation takes place deep in the Mott insulator regime with a large spectral gap (see also Fig.\ \ref{initial_state}). The code uses a fifth order Trotter decomposition and truncates the local occupation at 6 bosons. By performing an extensive scaling in Trotter step size as well as bond dimension, we have ensured that the simulation is stable with respect to these parameters and that the two parameters are chosen sufficiently small and large, respectively (Fig.\ \ref{fig:DMRG}). The simulation assumes open boundary conditions and the coherence length is evaluated by an exponential fit of the correlations at the centre of the system. In contrast to other settings where a sharp spatial transition in the decay behaviour was observed \cite{1308.4699}, the correlation decay is smooth for the ramps considered in this work (Fig.\ \ref{fig:numerics_exponential_decay}). For long $\tau_{\text{ramp}}$, the correlators are expected to decay exponentially only for large distances but to follow a power-law behaviour for short distances \cite{1308.4699}. However, we fit the decay with a pure exponential function as this describes the experimental results sufficiently well.

\begin{figure}[htb]
    \includegraphics[width=84mm]{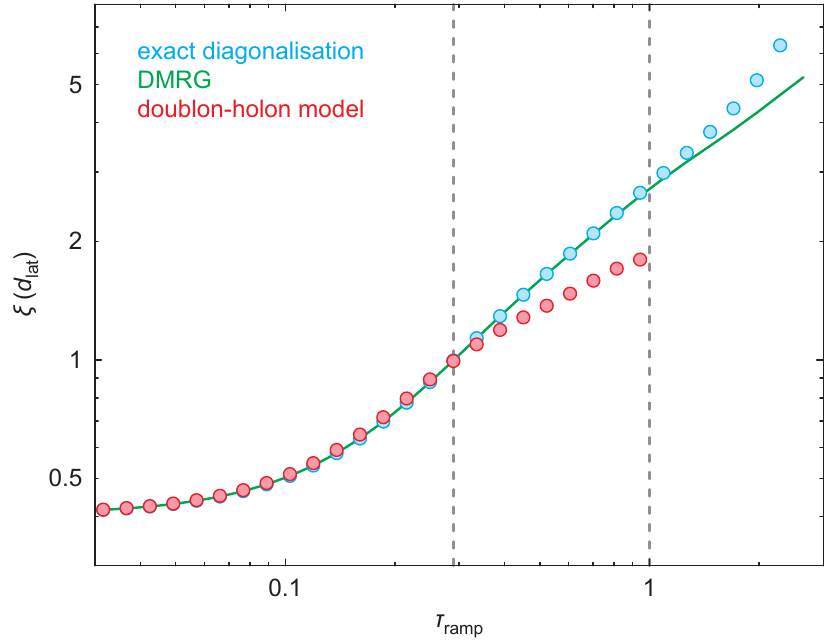}
    \caption[Numerical comparison]{Comparison between the various simulation methods for $(U/J)_{\text{f}}=2$ in 1D. The vertical dashed lines indicate the ramp times where the different methods start to deviate from each other.}
    \label{fig:numerical_comparison}
\end{figure}

The validity of the DMRG simulation was further cross-checked with the doublon-holon model (Supplementary section F), as well as an optimised exact diagonalisation code (Fig.\ \ref{fig:numerical_comparison}). The exact diagonalisation simulation is a Runge-Kutta numerical integration of a homogeneous Bose-Hubbard model on 15 sites with unity average filling, where the local occupation is truncated at 9 bosons. We assume periodic boundary conditions and all symmetries are taken into account to reduce the computational complexity. This comparatively small 1D system is able to capture the initial emergence of coherence of the large scale system considered in the experiment, because the relevant coherence lengths remain much smaller than the system size for the timescales at hand. Thus, the two simulations yield the same power-laws for the emergence of coherence.

\begin{figure}[htb]
    \includegraphics[width=84mm]{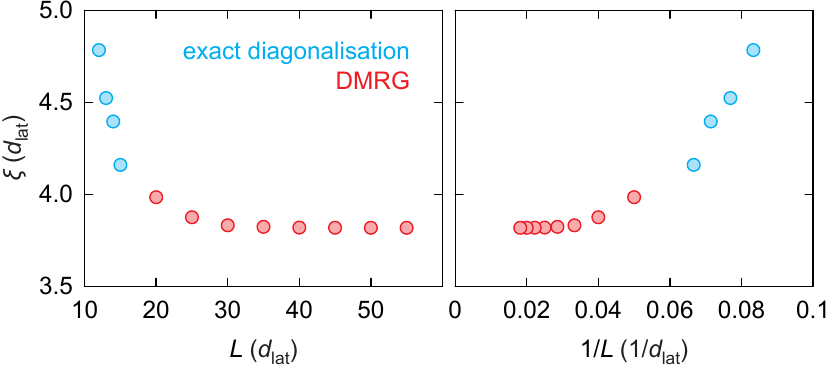}
    \caption[Finite size]{Finite size scaling for $(U/J)_{\text{f}}=1$ in 1D for a fixed ramp time $\tau_{\text{ramp}}=2.0$. Exact diagonalisation and DMRG data are plotted versus system size $L$ (left). A plot versus $1/L$ allows to extract the limit $L\rightarrow\infty$ (right).}
    \label{fig:finite_size}
\end{figure}

A deviation between DMRG and exact diagonalisation is only visible for longer simulation times, $\tau_{\mathrm{ramp}} \geq 1$. For those times, an extensive finite size scaling was performed (Fig.\ \ref{fig:finite_size}). As one would expect from light-cone-like arguments, for each ramp time, all systems larger than a particular size behave essentially like an infinite system.

\section{F: Quasiparticles and the doublon-holon model}
\label{dh}

The Doublon-Holon Fermionic Model (DHFM) offers an excellent approximation for the 1D Bose-Hubbard model in the strong interaction regime $U/J\gg 1$ with integer filling fraction $\bar n$. In this scenario, the ground state of the system is given by a Mott insulator with $\bar n$ particles per site. Due to the strong interaction coupling, the low energy subspace is well described by local occupations with values $\bar n-1$, $\bar n$, and $\bar n+1$. This suggests the introduction of two kinds of excitations on top of the background Mott insulator ground state, doublons and holons, corresponding to occupations $\bar n+1$ and $n-1$, respectively. The DHFM is still quantitatively valid in the regime of intermediate interactions, from the deep Mott insulator regime down to a value $(U/J)_\text{limit}\approx 8$ for $\bar{n} = 1$ \cite{Cheneau, Cheneau2012}. While the DHFM is unable to adequately describe the behaviour at the phase transition $(U/J)_\text{c}\approx 3.3$ ($\bar{n}=1$), it provides a very physical picture of the problem and has a broader parameter range than most methods. Furthermore, it is exactly solvable and therefore allows us to explore the evolution of the system for time-dependent parameters $U(t)$ and $J(t)$, even for several hundreds of bosons. Even though the modelled ramps cross the phase transition, we found that the DHFM reproduces the results of the full Bose-Hubbard model for sufficiently short ramp times.

\subsection{Formal description}

Following Refs.\ \cite{Cheneau, Cheneau2012}, we assume periodic boundary conditions and a filling of $ \bar{n} = 1 $ and start from the translation invariant Bose-Hubbard Hamiltonian
\begin{equation}
\label{eq:1d-boson-model}
H = \sum_{j=0}^{L-1}\Bigl[ -J \bigl(\hat a^{\dagger}_{j} \hat a_{j+1}+\text{h.c.}\bigr)+ \frac{U}{2}  \hat n_j (\hat n_j -1)  - \mu \hat n_j\Bigr],
\end{equation}
where $\hat n_j = \hat a^{\dagger}_{j} \hat a_{j}$, $L$ is the length of the chain and $\mu$ the chemical potential.

Deep in the Mott phase, local density fluctuations and occupation numbers are small. In this regime, a model in which the local Hilbert space dimension is truncated at a maximum of two bosons per site is expected to approximate the true dynamics well. With this truncation, we construct the new fermionic doublon ($\hat{d}$) and holon species ($\hat{h}$) via a double Jordan Wigner (JW) transformation. These operators represent localised excitations on top of a background state with a single particle per site. To solve the resulting model, we turn to the Fourier basis,
\begin{align}
  \label{eq:ham-decomposition}
  \begin{split}
    H=\sum_{k=0}^{L-1} H_k &= \sum_{k=0}^{L-1}  \left[\epsilon_{d} (k)  \hat {d}_k^\dagger  \hat {d}_k 
    +\epsilon_{h} (k)  \hat {h}_k^\dagger \hat h_k \right. \\
    &\left. \vphantom{\hat {h}_k^\dagger} + i  \sigma(k)(\hat {h}_{-k}^\dagger \hat d^\dagger_{k} +  \hat d_{-k}\hat  {h}_k ) -\mu \right],
  \end{split}
\end{align}
where
\begin{align}
\epsilon_h (k)&=\mu-2J \cos\left( \frac{2\pi}{L}k\right), \;
\epsilon_d (k)=2 \epsilon_h+U-3\mu,\\
\sigma(k)&=2\sqrt{2} J \sin\left( \frac{2\pi}{L}k\right).
\end{align}
Eq.\ \eqref{eq:ham-decomposition} incorporates the additional approximation of unconstrained fermions: A projection term eliminating the unphysical situation of having a doublon and a holon at the same site has been dropped. In the deep Mott insulator regime, this approximation holds to excellent accuracy, as two excitations at the same site are very unlikely due to the low density of excitations.

With Eq.\ \eqref{eq:ham-decomposition}, we can model the experimental ramp by introducing time-dependent parameters $U(t)$, $J(t)$. For a given time, we can always exactly diagonalise the resulting Hamiltonian in this approximation by making use of a Bogoliubov transformation into new modes (quasiparticles with definite momenta, $\gamma_{\hat d,k}$, $\gamma_{\hat h,-k}$). From the dispersion relation, we obtain, in the same fashion as in Ref.\ \cite{Cheneau2012}, a maximum velocity for the spread of correlations that is reminiscent of a Lieb-Robinson bound. This velocity is the maximum relative velocity of pairs of quasiparticles of distinct types and opposite momenta, which are created simultaneously during the quench,
\begin{align}
  \label{eq:maximum-speed}
  \mathcal{V} = \max_k \left|v_{\gamma_{\hat h,{-k}}}\! \!- v_{\gamma_{\hat d,k}}\right| 
  = 6J-\frac{96}{16} \frac{J^3}{U^2} \! + O \left( \frac{J^4}{U^3}\right).
\end{align}
To obtain the correlators needed for the comparison with the experimentally measured TOF images, we use the fact that the time dependence does not affect the decomposition of the system into a tensor product of independent modes labelled by momentum index $k$ (Eq.\ \eqref{eq:ham-decomposition}). As the ramp starts in the ground state of the deep Mott insulator, the initial state can be well approximated by a product state with one particle per site and thus without any doublons or holons: $\hat d_k |\psi_0\rangle \!= \! \hat h_k |\psi_0\rangle \! = \!0$. Therefore, we can work separately for each $k$ in the subspace spanned by $\ket{0}_k:=\ket{0}_{\hat  d_{k},\hat {h}_{-k}}$ and $\ket{1}_k:=\hat  d^\dagger_{k}\hat {h}^\dagger_{-k}\ket{0}_k$. The time-evolved state for such a time-dependent Hamiltonian can be written as $\ket{\psi(t)}=\bigotimes_k \ket{\psi(t)}_k $, with
\begin{align}
\ket{\psi(t)}_k&=C_0 (k, t)\ket{0}_k+C_1(k,t)\ket{1}_k
\end{align}
and the initial state is described by $C_0(k,0)=1$ and $C_1(k,0)=0$ for all $k$. The time evolution is then given by a system of coupled differential equations for $C_0(k,t)$ and $C_1(k,t)$.

\subsection{Correlator techniques}

By solving the corresponding equations, we obtain the mode decomposition of a state at any time after a quench starting deep in the Mott phase, as long as we work within the validity regime of the method. To subsequently obtain the original bosonic two-point correlators $\langle  a_\mu^\dagger a_\nu \rangle$ is challenging, since it involves the computation of many-point fermionic correlators,
\begin{align}
\label{eq:correlator_calc}
\langle \hat a^\d_\mu \hat a_\nu \rangle =  2 \langle \hat d^\d_\mu \hat Z^{\dagger (d)}_\mu \hat Z^{(d)}_\nu \hat d_\nu \rangle +  \sqrt{2} \langle \hat Z^{(h)}_\mu \hat h_\mu \hat Z^{(d)}_\nu \hat d_\nu\rangle \\
+ \sqrt{2} \langle  \hat d^\d_\mu \hat Z^{\dagger (d)}_\mu \hat h^\dagger_\nu \hat Z^{\dagger (h)}_\nu \rangle + \langle \hat Z^{(h)}_\mu \hat h_\mu \hat h^\dagger_\nu \hat Z^{\dagger (h)}_\nu \rangle,
\end{align}
where $\hat Z$ is the string operator corresponding to the JW transformation. In the calculation, each of these correlators is expanded using auxiliary Majorana fermions and then decomposed, by applying Wick's theorem, into a Pfaffian involving products of two-point fermionic correlators. These two-point correlators can be calculated by relating them to the coefficients $C_0(k,t)$, $C_1(k,t)$. The whole derivation is rather extensive and not included here.

\subsection{Applicability of the model}

To assess the validity of the DHFM for our setting, we compared the results for a quench in the Bose-Hubbard model for the DHFM with 64 sites with the exact diagonalisation routine for 15 sites. Although we only show the results for the quench $(U/J)_{\text{i}}=47$, $(U/J)_{\text{f}}=2$, we have compared the plots for all other ramps and checked that, qualitatively, they all share the same behaviour.

We can easily verify that for short ramp times the DHFM follows accurately the exact diagonalisation numerics, even though the ramp reaches into the superfluid regime, where the model is not applicable anymore (Fig.\ \ref{fig:numerical_comparison}). For fast quenches, few doublons and holons are created and do not have enough time to propagate. Thus, the limitations of the model, i.e.\ less than three particles per site and unconstrained fermions, are not reached. For slow quenches, however, this statement does not hold anymore and the prediction deviates from the exact treatment.

\subsection{Lieb-Robinson bounds and coherence length growth}

\begin{figure}[htb]
  \includegraphics[width=84mm]{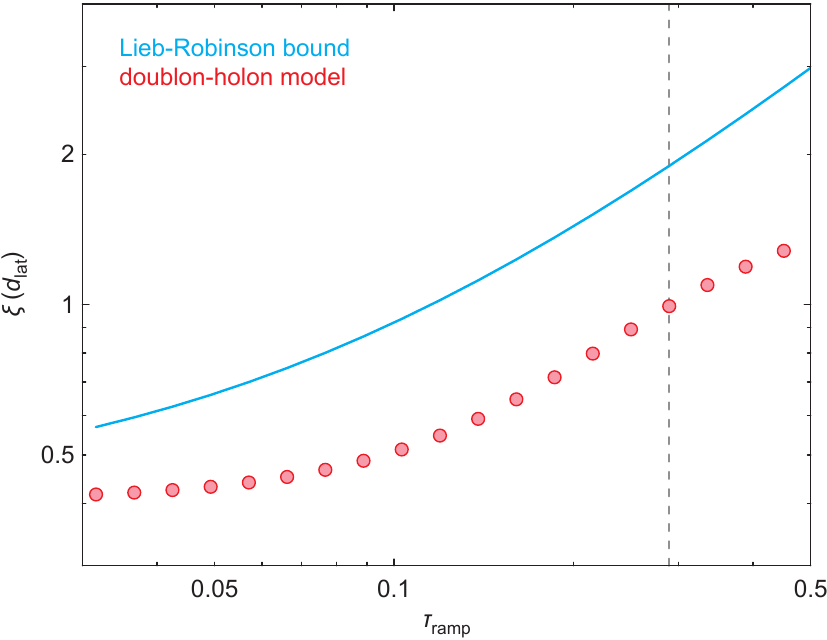}
  \caption{Correlation length of the DHFM and the Lieb-Robinson bound provided by the group velocity of the excitations for $(U/J)_{\text{f}}=3$ in 1D. The vertical dashed line indicates the ramp time for which deviations between the doublon-holon model and the full exact diagonalisation appear (Fig.\ \ref{fig:numerical_comparison}).}
  \label{fig:LRb}
\end{figure}

To develop a better intuition of the underlying physical processes, we connect our results with Lieb-Robinson (LR) bounds. As rigorous Lieb-Robinson bounds are not applicable to strongly correlated Bose-Hubbard models, we consider the time-dependent maximum velocity $\mathcal{V}$ for the spread of quasiparticle pairs introduced in Eq.\ \eqref{eq:maximum-speed}. By the end of the quench, significant spreading must be less than $\int_0^{\tau_{\text{ramp}}} \text{d}t\,\mathcal{V}(t)$, yielding a crude upper bound for the coherence length 
\begin{equation}
  \xi(\tau) \leq \xi_0 + \int_0^{\tau_{\text{ramp}}} \text{d}t\,\mathcal{V}(t).
	\label{eq:lr-bound}
\end{equation} 
This inequality considers spreading of quasiparticles and an associated increase of the coherence length, even when the system is deep in the Mott insulating regime and close to the ground state, where excitations are absent. Nevertheless, it is still interesting to see that this approximation, which captures most of the intuitions underlying the DHFM, already shows a resemblance with the exact computation (Fig.\ \ref{fig:LRb}).

\section{G: Kibble-Zurek mechanism}
The Kibble Zurek mechanism (KZM) provides an intuitive explanation of the defect formation and the resulting coherence lengths when a second order phase transition is crossed at a finite velocity. In the quantum case, the situation can usually be captured in the following way:
\begin{itemize}
  \item Initially, the system is at equilibrium in a disordered phase.
  \item By changing a parameter of the Hamiltonian, the system is driven at a certain velocity across critical lines towards an ordered phase that breaks the symmetry.
  \item As the transition is crossed within finite time, suf\-fi\-cient\-ly remote regions are causally disconnected. Hence, the symmetry cannot be broken homogeneously and the local order parameter takes different values in different spatial domains, giving rise to defects.
\end{itemize}
The KZM provides a simple argument to estimate the size of the domains as well as the final correlation length. The essence of the KZM states that key properties of the final state scale like power-laws with the quench time and the exponent depends only on the critical scaling of the model at equilibrium. Thus, the KZM can be seen as an extension of the universality of the equilibrium features of a model to its dynamics. The KZM has successfully been applied to many experimental results and numerical simulations \cite{ZurekDornerZoller,cherng_levitov,pekker_demler,Ions,tzker_Plenio_Zurek_et_al__2013,Dziarmaga}.

\subsection*{Adiabatic -- frozen evolution}
Several ground state properties near criticality can be captured by means of critical exponents. More concretely, the spectral gap $\Delta>0$ and the correlation length $\xi>0$ of the system are expected to scale with the distance $|\lambda-\lambda_\text{c}|$ from the critical point $\lambda_\text{c}$ as
\begin{align}
\label{eq:equilibration-time}
\Delta &\propto |\lambda-\lambda_\text{c}|^{z\nu}\,\,\,\text{and}\\
\xi &\propto |\lambda-\lambda_\text{c}|^\nu\, , 
\label{eq:corr-length}
\end{align}
where $\nu,z>0$ are the critical exponents and $\lambda$ is the control parameter of the Hamiltonian. The scaling law emerging from the KZM is typically stated in a narrative that suggests to divide the evolution into an \emph{adiabatic} and a \emph{frozen} regime. 
\begin{itemize}
  \item \emph{Adiabatic dynamics}: The system is initially in the ground state protected by a spectral gap $\Delta>0$. As long as the gap is large compared to the change rate of the control parameter, typically estimated by $ \dot \Delta\ll \Delta^2$, the adiabatic approximation holds: At all times, the state of the system is well approximated by the respective ground state.
  \item \emph{Adiabaticity breaking}: As the control parameter is driven at a finite velocity $v=\dot\lambda$ and the gap closes to zero at the critical point, at some moment ($\lambda=\lambda_\text{Z}$), the system cannot follow the change of parameters anymore and the evolution fails to be adiabatic. If this occurs sufficiently close to the phase transition and if the ground state properties are meaningfully captured by the critical exponents defined above, the adiabaticity condition, together with Eq.\ \eqref{eq:equilibration-time}, implies a scaling relation for the distance between the breakdown of adiabaticity and the critical point,
\begin{equation}
|\lambda_\text{Z}-\lambda_\text{c}|\propto v^{-{1}/{(1+\nu z)}}\, .
\end{equation}
\item \emph{Frozen dynamics}: From this moment on, the dynamics are considered `frozen' in the sense that long-range correlations can no longer be established and the final value of the coherence length of the system is determined by Eq.\ \eqref{eq:corr-length},
\begin{equation}
\xi_{\textrm{final}}\propto v^{-{\nu}/{(1+\nu z)}}\, .
\end{equation}
This frozen correlation length can also be connected with a `sonic horizon' \cite{1310.1600}.
\end{itemize}

\subsection*{Validity of the adiabatic -- sudden approximation}
\begin{figure}[htb]
  \centering
  \includegraphics[width=84mm]{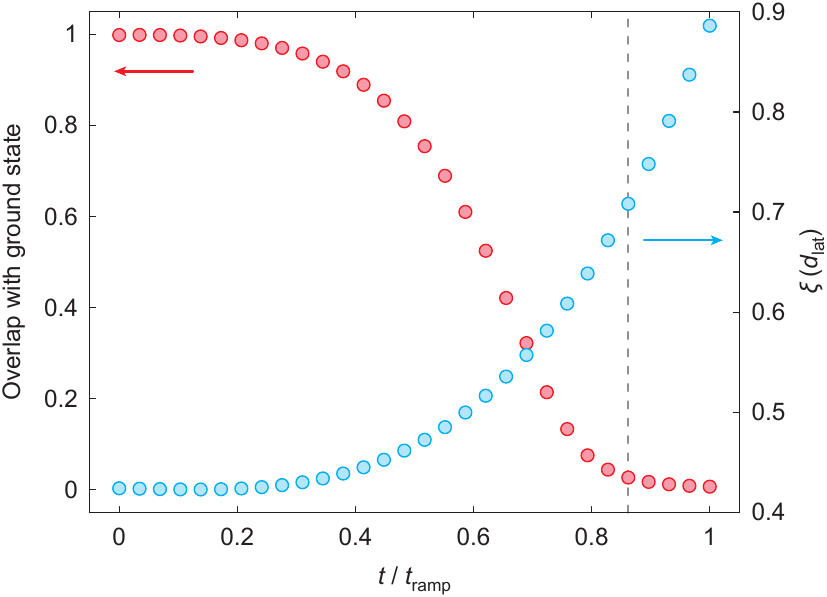}
  \caption[no freezing]{Time evolution for $(U/J)_{\text{i}}=47$, $(U/J)_{\text{f}}=2$ in 1D for a fixed ramp time $\tau_{\text{ramp}}=0.25$. Red dots represent the overlap with the ground state. Blue dots display a change of the coherence length $\xi$. The vertical dashed line indicates the phase transition.}
  \label{no_freezing}
\end{figure}

Both the simplicity and the predictive power of the previous argument are striking. It has been the subject of a large body of literature, how the KZM could actually be realised and rigorously true in strongly correlated models, and several questions are still not quite satisfactorily resolved. Specifically in the situation at hand, the above mindset appears to be not directly applicable. The time evolution is, within excellent accuracy, indeed adiabatic far away from the phase transition (Fig.\ \ref{no_freezing}). When the gap becomes small compared to the change of the Hamiltonian, adiabaticity clearly breaks down. It is very unclear, however, in what sense the correlation dynamics could be conceived as `frozen'. In fact, in the Bose-Hubbard model discussed here, the bulk of the dynamics happens just around the critical point (Fig.\ \ref{no_freezing}). In standard text books \cite{Messiah}, one can find rigorous lower bounds for the quench rate of the Hamiltonian for which the full system remains unchanged. Yet, these bounds are very restrictive and strongly depend on the system size. These limitations, however, are not an artefact of the bounds: A vast range of dynamical processes exists which are neither adiabatic nor sudden.

\subsection*{Approaches to the study of the dynamics of phase transitions}
Exactly solvable integrable models provide a reasonable guideline to investigate how faithfully the KZM describes slow quenches across second order phase transitions. This approach has been pursued specifically for translational invariant free fermionic systems, where the Hilbert space decouples into subspaces corresponding to different independent modes. The transition probability for a given two-level mode can be estimated via the Landau-Zener formula. In this scenario, the size of the domains is determined by the smallest wavelength among all excited modes and can be proven to scale as a power-law with the quench time, with the exponent provided by the critical exponents \cite{DziarmagaReview,0908.2922}. Via adiabatic perturbation theory \cite{ngupta_Silva_Vengalattore_2011,0910.3692}, which follows a similar approach despite being mathematically inequivalent, power-law scalings have also been derived for more complex and non-integrable models. There is increasing evidence that power-laws reminiscent of the KZM emerge from a deeper scale invariance in the system \cite{sondhi_scaling,Kolodrubetz,1311.1543}. This intuition can again be corroborated by free bosonic or fermionic models.

\subsection*{Applicability of the KZM to the Bose-Hubbard model}

Fig.\ \ref{experiment_exact_diagonalisation} shows that the complexity of the dynamics of the Bose-Hubbard model, when quenched from the Mott insulator to the superfluid phase, is not satisfactorily captured by the KZM in the considered regime of fast and intermediate ramps. Although for intermediate ramp times, the coherence length follows a power-law, the KZM prediction turns out to be too crude to model the full complexity of this transition: The observed exponents strongly depend on the final position of the quench within the superfluid phase and depend much less on dimensionality than suggested (Fig.\ \ref{pos_T_neg_T_and_dimensions}).

There are several plausible reasons why KZM is not sufficient to capture the observed dynamics. A number of variants of the KZM as well as corrections have recently been proposed that take into account several specifics of slow quenches. These include the initial and final values of the control parameter \cite{1001.0693, sondhi_scaling}, finite size effects \cite{ZurekDornerZoller,ovi_Rossini_Fazio_Santoro_2009}, particularities of the quench schedule chosen \cite{Sen_Sengupta_Mondal_2008,0908.2922,haque,Quan_Zurek_2010}, and multi-critical points as well as the dimensionality of the critical surface \cite{karan_Mukherjee_Dutta_Sen_2009,Divakaran_Dutta_Sen_2008,deng_viola,polkovnikov_universal}. Furthermore, it was pointed out recently that the 1D case requires a more careful analysis due to its Kosterlitz-Thouless transition \cite{1312.5139}, characterised by an exponential closing of the gap. In the following, we investigate the relevance of these issues to this work.

\begin{figure}[htb]
  \centering
  \includegraphics[width=84mm]{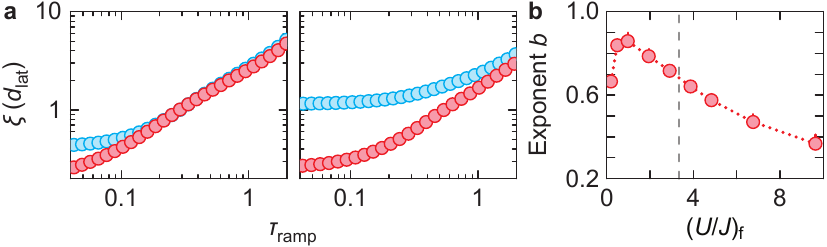}
  \caption[deep ramps]{\textbf{a}, Comparison of the 1D behaviour between the experimentally considered ramp (blue) and one that starts in a deeper lattice of $V_{\text{lat}}=45E_{\text{r}}$ (red) for $(U/J)_{\text{f}}=2$ (left) and $(U/J)_{\text{f}}=0.5$ (right). Data is obtained with exact diagonalisation on 15 sites. \textbf{b}, Power-law exponents from exact diagonalisation when starting the quenches in a deeper lattice. The dotted line guides the eye and the vertical dashed line indicates $(U/J)_{\text{c}}$.}
  \label{deeper_ramps}
\end{figure}

We have verified that the influence of the starting point of the ramp is marginal for most quenches considered in this work. Merely for the quench that proceeds deepest into the superfluid regime, an analogous quench starting in a deeper lattice of $V_{\text{lat}}=45E_{\text{r}}$ alters the power-law exponent. Yet, the characteristic behaviour of a decreasing power-law exponent for smaller $(U/J)_{\text{f}}$ is still present (Fig.\ \ref{deeper_ramps}).

The small resulting coherence lengths allow us also to rule out finite size effects as the origin for our observations. Numerical simulations also support this as they agree both among themselves and with the experiment in a range of system sizes from 12 to 70 sites.

In the 1D case, the Mott to superfluid transition is of a Kosterlitz-Thouless type and hence the quench schedule is usually claimed irrelevant. This might, however, deserve further attention in the light of the newest results on the dynamics of such transitions \cite{1312.5139}. In higher dimensions, the observed power-law could in principle be altered by the quench schedule, but the experimental ramp is close to linear in a large area around the phase transition at $(U/J)_{\text{c}}=16.74$ \cite{PhysRevA.77.015602} in 2D and $29.36$ \cite{PhysRevB.75.134302} in 3D, such that this influence is expected to be small (Fig.\ \ref{eoc_sequence}b).

\begin{figure}[htb]
  \centering
  \includegraphics[width=84mm]{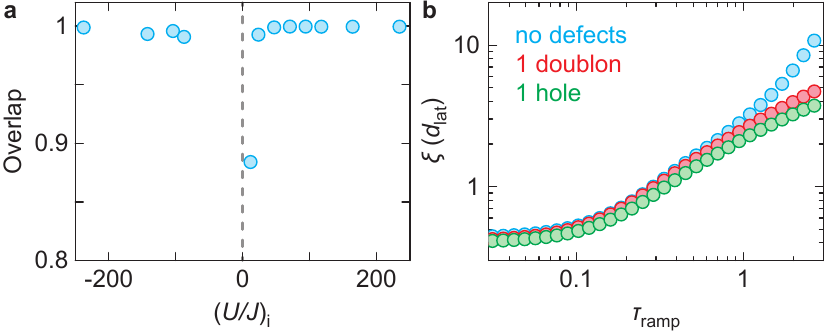}
  \caption{\textbf{a}, Adiabaticity of the Feshbach ramp for positive and negative temperatures. For positive $(U/J)_{\text{i}}$, this refers to the overlap with the ground state. In the case of attractive interactions, corresponding to negative temperatures, the overlap with the highest excited state is plotted. \textbf{b}, Influence of defects in the initial state obtained by an exact diagonalisation simulation on 12 sites.}
  \label{initial_state}
\end{figure}

The potential influence of the initial state has been studied much less in the literature \cite{deng_viola_2011}. We have verified numerically that the Fesh\-bach ramp of the 1D experiment indeed prepares the ground state of the system, as presumed by the typical Kibble-Zurek setting (Fig.\ \ref{initial_state}a).

To investigate the influence of defects and finite temperature effects, we ran exact diagonalisation simulations on a small system where the initial state involves a doublon or a hole. This leads to a slight decrease of the exponents compared to the zero temperature case (Fig.\ \ref{initial_state}b). It seems unlikely that this could satisfactorily explain the unexpected role of dimensionality (Fig.\ \ref{pos_T_neg_T_and_dimensions}).

Sufficiently slow ramps are another usually assumed condition for the KZM, even though it is difficult to give an estimate for this limit. It seems reasonable to demand that adiabaticity should break down only sufficiently close to the phase transition, such that knowledge of the ground state scaling is sufficient to describe the time-evolved state at that point. It remains an interesting challenge to obtain quantitative estimates for this condition that do not require the full knowledge of the time evolution. A complete formulation of the KZM should thus also incorporate a range of quench times for which it is expected to approximate the true scaling laws well. Presumably, one cannot expect the range to be universal and only depend on the critical exponents, but are there guidelines independent of very specific features of the Hamiltonian that may give rise to good bounds? To what extent would the initial state and the precise quench schedule affect it?

It thus remains an intriguing and general question how much knowledge about the performed quench and the system is necessary for an accurate characterisation of the evolution. While obviously a full knowledge of the quench, the energy levels and the eigenstates of the model is sufficient, the success of the Kibble-Zurek mechanism suggests that a lot less knowledge is typically necessary. A satisfactory answer to this question will be crucial for a theory of the dynamics of quantum phase transitions. The present work, presenting reliable data both from numerical and quantum simulations, is an invitation in this direction.

\vspace{5cm}

\end{document}